\documentclass[twocolumn]{aastex63}
\usepackage{mathtools}
\usepackage{xspace}
\usepackage{dsfont}
\usepackage[nolist]{acronym} 
\usepackage[nameinlink,capitalise,noabbrev]{cleveref}

\newcommand{\REV}[1]{{{#1}}}
\newcommand{\RREV}[1]{{{#1}}}

\makeatletter
\usepackage{etoolbox}
\patchcmd\H@refstepcounter{\protected@edef}{\protected@xdef}{}{}
\makeatother

\defcitealias{moon24}{Paper~I}
\defcitealias{paperII}{Paper~II}

\newcommand{\secresolution}{3.3\xspace}
\newcommand{\seccoretracking}{3.5\xspace}
\newcommand{\secresolutiondiscussion}{5.2\xspace}
\newcommand{\seccontrolledcollapse}{5.3\xspace}

\newcommand{\eqsigmaoneddef}{(14)\xspace}
\newcommand{\eqlambdastors}{(16)\xspace}

\newcommand{\eqlageom}{(61)\xspace}

\newcommand{\eqnminsonic}{(51)\xspace}

\newcommand{\figcoreevolution}{11\xspace}
\newcommand{\figpredictedforceevolution}{14\xspace}
\newcommand{\figlinewidthsize}{15\xspace}
\newcommand{\figrsrhocevolution}{16\xspace}

\shorttitle{Critical Core Properties}
\shortauthors{Moon \& Ostriker}

\begin{document}

\title{Prestellar Cores in Turbulent Clouds: Properties of Critical Cores}

\author[0000-0002-6302-0485]{Sanghyuk Moon}
\affiliation{Department of Astrophysical Sciences, Princeton University,
  Princeton, NJ 08544, USA}
\author[0000-0002-0509-9113]{Eve C.\ Ostriker}
\affiliation{Department of Astrophysical Sciences, Princeton University,
  Princeton, NJ 08544, USA}
\email{sanghyuk.moon@princeton.edu, eco@astro.princeton.edu}

\begin{acronym}
    \acro{BE}{Bonnor-Ebert}
    \acro{TES}{turbulent equilibrium sphere}
    \acro{ISM}{interstellar medium}
    \acro{MHD}{magnetohydrodynamic}
    \acro{GMC}{giant molecular cloud}
    \acro{VL2}{second-order van Leer}
    \acro{LP}{Larson-Penston}
    \acro{CMF}{core mass function}
    \acro{CCMF}{``critical core mass function''}
    \acro{SMF}{sink particle mass function}
    \acro{IMF}{initial mass function}
    \acro{PDF}{probability distribution function}
    \acro{YSO}{young stellar object}
    \acro{RMS}{root mean square}
    \acro{AMR}{adaptive mesh refinement}
    \acro{SFE}{star formation efficiency}
\end{acronym}

\begin{abstract}
A fraction of the dense cores within a turbulent molecular cloud will eventually collapse to form stars.
Identifying the physical criteria for instability and analyzing critical core properties is therefore necessary to star formation theory.
Here we quantify the characteristics of an ensemble of ``critical cores'' on the verge of collapse.
This critical epoch was identified in a companion paper, which followed the dynamical evolution of prestellar cores in numerical simulations of turbulent, self-gravitating clouds.
We find that radial profiles of density and turbulent velocity dispersion in individual critical cores are consistent with our new model for turbulent equilibrium spheres (TESs).
While a global linewidth--size relation exists for a cloud with given size and Mach number, the turbulent scaling relations around each core exhibit significant variations.
As a result, there is no single density threshold for collapse; instead, cores collapse at a range of densities determined by the local sonic scale and gravitational potential environment.
The critical cores in our simulations are mostly transonic; we do not find either purely thermal or highly turbulent cores.
\REV{In our low Mach number model which better resolves the characteristic mass and sonic scales, we find marginal evidence that the core mass function (CMF) of critical cores peaks around a characteristic mass scale associated with the large-scale cloud properties.}
We highlight the importance of constructing the CMF at the critical time \REV{for the purpose of testing  gravoturbulent fragmentation theories}, and derive the resolution requirements to unambiguously identify the peak of the CMF.
\end{abstract}

\section{Introduction}\label{sec:intro}

Dense cores are transient structures within turbulent \acp{GMC}, with their properties continuously changing in space and time \citep[see][for related reviews]{bergin07,difrancesco07,mckee07,andre14,offner14,padoan14}.
While the physical properties of \emph{prestellar} cores --- which ultimately collapse, leading to star formation --- are of great interest, it has not been clear exactly what the criterion is for a given core to become unstable and therefore unambiguously 
``prestellar.''
In observational surveys, the simplest working definition for prestellar candidates is cores that appear gravitationally bound, based on estimates of their internal energy, although more detailed criteria have also been applied \citep[e.g.][]{andre2000,konyves15}.
However, simulations that trace the evolution of dense structures show that many stochastically cycle through different phases (due to their turbulent environment), rather than following a single evolutionary sequence, and in particular cores with ``prestellar'' characteristics sometimes subsequently disperse \citep{offner22}.

Notwithstanding the ambiguities in recognizing prestellar cores, their very nature --- objects that will collapse to form a star or stellar system in the near future --- provides at least two physically meaningful common reference points in time shared across all cores: (1) The time when they \emph{initiate} gravitational runaway (hereafter ``critical time'') and (2) the instant they \emph{complete} it by forming a nascent protostar at the center (``collapse time'').
While these milestones would be essentially impossible to pinpoint in observations, they can be identified on an individual core basis within numerical simulations of gravo-turbulent fragmentation.

The second of the two above events is easier to pick out in simulations, since this is when runaway collapse leads to a $\rho \propto r^{-2}$ singular density profile \citep{larson69,penston69}.\footnote{Strictly speaking, it is only at the center of this power-law profile that a protostar-disk system begins to grow.} In particular, in numerical simulations where collapsing centers are replaced with sink particles, the collapse time of each core can be identified as the moment when a sink particle is created.
For example, \citet{gong15}, \citet{padoan20}, and \citet{pelkonen21} used the closest snapshot to the creation time of each sink particle to define their ``progenitor core'' as a gravitationally bound region around the site where the sink is created.

By measuring the core mass at each core's collapse time, \citet{gong15}  demonstrated that the \ac{CMF} has a well-defined peak which converges with increasing numerical resolution.
\citet{pelkonen21}, taking a slightly different approach, varied the resolution for the clumpfind algorithm rather than the underlying simulation resolution, and also found convergence in the \ac{CMF} at the time of collapse.
In the majority of numerical studies of gravo-turbulent fragmentation, however, it is not the \ac{CMF} at a characteristic epoch for each core or sink's evolution that is measured, but instead the \ac{SMF} at an arbitrarily selected time common to all particles \citep[e.g.,][]{haugbolle18,lee18II,guszejnov20}.
The resulting \ac{SMF} has not shown numerically converging behavior, unlike the \ac{CMF} based on individual collapse times \citep[see, however,][]{haugbolle18,pelkonen21}.
As we shall discuss, the reason for the difference between \ac{CMF} convergence and \ac{SMF} non-convergence likely relates to fragmentation in late stages of evolution.
But in any case, in order to test theories of turbulent fragmentation that are the basis for models of the initial mass function \citep[e.g.,][]{padoan02,hennebelle08,hopkins12}, it is necessary to construct the \ac{CMF} itself at characteristic epochs of evolution.

While the collapse time provides a convenient reference point and can be easily identified in numerical simulations, it is the \emph{critical time} that is most relevant for the point of comparison to theories.
\citet{collins24}, by analyzing the collapse histories of simulated cores via tracer particles, found that there exists a characteristic epoch after which mass is rapidly delivered to the core center, raising the core density in a runaway fashion.
Based on this observation they defined a proxy for the critical time (termed ``singularity time'' in their work) by choosing a threshold on the density time derivative.
They found that, as each core approaches its critical time, the radial density profiles of cores develop a characteristic core--envelope structure while the mean radial and tangential velocities exhibit a common behavior of increasing in magnitude with increasing radius.
The identification of this characteristic epoch reveals a common physical process occurring in an ensemble of cores that is otherwise blurred by evolution.
However, the question of what determines this critical time still remains to be answered.

Related to the critical time is a common narrative for the evolution of prestellar cores, where certain critical conditions trigger gravitational runaway (\citealt{mckee07}; \citealt{andre14}; \citealt{offner14}; \citealt{padoan14}).
Among others, the \ac{BE} sphere and its stability property \citep{ebert55,bonnor56,ebert57} have often been used as a theoretical tool for determining critical conditions \citep[e.g.,][]{krumholz05,padoan11}.
However, the applicability of the \ac{BE} model to cores within turbulent molecular clouds is limited, because (1) real cores exist as a part of a turbulent continuum unlike an isolated \ac{BE} sphere truncated by ``external pressure,'' and (2) the model does not account for internal turbulent velocities expected for cores forming within a turbulent cloud.

In a series of papers, we have begun a comprehensive analysis of prestellar core evolution to address the limitations pertaining to the \ac{BE} model and to answer the question of exactly what triggers the onset of collapse.
In \citet[hereafter \citetalias{moon24}]{moon24}, we developed a new theoretical model for equilibrium spherically-symmetric cores supported by both thermal and turbulent pressure, with the solutions obtained by directly solving the time-steady, angle-averaged equations of hydrodynamics.
A distinguishing feature of the resulting family of solutions, which we term the \ac{TES}, is that the turbulent pressure naturally arises from a power-law velocity structure function rather than from a phenomenological equation of state.
The \ac{TES} model also enjoys a useful behavior of reducing to the \ac{BE} model in the limit of vanishing turbulent velocities.
\citetalias{moon24} found that, similar to the case for \ac{BE} spheres, solutions can be classified as stable or unstable to radial perturbations.
For a given set of turbulence parameters and central density, cores exceeding a certain critical size are unstable.
The radius, mass, and center-to-edge density contrast of critical \acp{TES} all increase with increasing turbulent velocity dispersion (or equivalently, decreasing sonic scale).
This implies that a nascent core forming in a highly turbulent region would initially be in the stable, subcritical regime.

Based on the stability properties of \acp{TES} and from a general consideration of tidal gravitational fields in core-forming regions, \citetalias{moon24} proposed an evolutionary scenario for ``successful'' prestellar cores formed by converging turbulent flows.
In this scenario, nascent cores evolve in the direction of increasing density (as mass is added) and decreasing turbulence (as dissipation progresses).
When converging flows are strong enough or turbulence sufficiently dissipates, the critical radius $r_\mathrm{crit}$ moves inside the ``tidal radius'' $r_\mathrm{tidal}$ set by the landscape of the gravitational potential, and a core becomes unstable and collapses.
Otherwise, it remains stable and is eventually dispersed by turbulence (see Figure 11 of \citetalias{moon24} for a schematic illustration for this scenario).
To test this scenario and identify the critical conditions for collapse, in \citet{paperII} (hereafter \citetalias{paperII}) we conducted and carefully analyzed a suite of numerical simulations of turbulent, self-gravitating clouds.
\citetalias{paperII} presents results from a comprehensive analysis of each core's dynamical evolution from its formation, through the onset of collapse, to the point when the central density blows up, signaling the beginning of the protostellar phase.
By measuring radial forces acting within the cores over time, \citetalias{paperII} directly identified the critical time when the net radial force becomes negative, instigating runaway collapse.
We found good agreement between the empirically identified critical time and the epoch when a core is predicted to be unstable according to the \ac{TES} model, thus providing an answer as to what determines the critical time.

In this paper, we analyze the simulation results of \citetalias{paperII} at the critical time of each core to investigate the physical properties of prestellar cores at the onset of collapse.
We shall show that the cores defined at their respective critical time (referred to as ``critical cores'' throughout this work) are mostly transonic in terms of their internal turbulence (turbulent Mach number $\sim 0.5\text{--}1.5$), with their structure overall consistent with the \ac{TES} model.
We shall also show that the local turbulent scaling relations within each core exhibit significant variations above and below the average linewidth--size relation for the entire cloud, with higher density in more turbulent critical cores.
As a result, there is a wide range of critical densities for collapse, rather than a single threshold density.
We shall also present statistical distributions of core masses, sizes, densities, durations of different evolutionary stages, and accretion rates.

The remainder of this paper is organized as follows.
In \cref{sec:definitions}, we define key physical quantities related to our \ac{TES} model and numerical simulations.
In \cref{sec:critical_core_properties}, we first (\cref{sec:radprof}) present the radial density and velocity dispersion profiles of critical cores and compare them with the \ac{TES} models.
We then (\cref{sec:linewidth_size}) show that the locally constructed linewidth--size relations exhibit significant variations around the mean relation, causing cores to become unstable at a wide range of densities (\cref{sec:core_properties}).
We also present probability distributions of various physical quantities measured for critical cores, including the \ac{CMF} (\cref{sec:CMF}).
\cref{sec:accretion} presents the evolution of the mass inflow rates and, by using them, characterizes evolutionary timescales at each stage.
We discuss implications of our results in \cref{sec:discussion} and conclude in \cref{sec:conclusion}.

\section{Key Definitions}\label{sec:definitions}

In this section, we provide key concepts and definitions of physical quantities that will be used throughout the paper.
We refer the reader to \citetalias{moon24} and \citetalias{paperII} for complete descriptions of our \ac{TES} model and numerical simulations, respectively.

\subsection{TES Model}

The radius and mass scales most relevant to isothermal self-gravitating objects supported entirely by thermal pressure are the critical \ac{BE} radius and mass given by
\begin{equation}\label{eq:rbe}
  R_\mathrm{BE}(\overline{\rho}) = 0.762 \frac{c_s}{G^{1/2}\overline{\rho}^{1/2}},
\end{equation}
\begin{equation}\label{eq:mbe}
    M_\mathrm{BE}(\overline{\rho})\equiv \frac{4\pi}{3} \overline{\rho}R_\mathrm{BE}^3(\overline{\rho}) = 1.86 \frac{c_s^3}{G^{3/2}\overline{\rho}^{1/2}},
\end{equation}
where $c_s = (kT/\mu)^{1/2}$ is the isothermal sound speed with temperature $T$ and mean molecular weight $\mu$, 
$G$ is the gravitational constant, and $\overline{\rho}$ is the average density within a core.

\citetalias{moon24} generalized these results for \acp{TES} characterized by the power-law linewidth--size relation of the form
\begin{equation}\label{eq:linewidth_size}
    \left< \delta v_r^2\right>_\rho^{1/2} = c_s\left(\frac{r}{r_s}\right)^p = c_s\left(\frac{\xi}{\xi_s}\right)^p,
\end{equation}
where $\delta v_r \equiv v_r - \left< v_r \right>_\rho$ is the radial velocity fluctuation (meridional, $\delta v_\theta$, and azimuthal, $\delta v_\phi$, velocity fluctuations are similarly defined), $r_s$ is the sonic radius, and $p$ is the power-law index.
In the second equality,
\begin{equation}\label{eq:xi_def}
    \xi \equiv \frac{\sqrt{4\pi G \rho_c} r}{c_s}
\end{equation}
is the dimensionless radius normalized with respect to the central density $\rho_c$, which also defines the dimensionless sonic radius $\xi_s$ where $r = r_s$.
The angled bracket in \cref{eq:linewidth_size} is a mass-weighted angle-averaging operator defined by
\begin{equation}\label{eq:mass-weighted-angle-average}
  \left<Q \right>_\rho \equiv \frac{\oint_{4\pi} \rho Q\,d\Omega}{\oint_{4\pi} \rho\,d\Omega},
\end{equation}
where $\rho$ is gas density and $Q$ is a physical quantity to be averaged.
It is related to the volume-weighted average
\begin{equation}\label{eq:volume-weighted-angle-average}
  \left<Q \right> \equiv \frac{1}{4\pi} \oint_{4\pi} Q\, d\Omega
\end{equation}
by $\left<Q \right>_\rho \equiv \left<\rho Q \right>/ \left<\rho \right>$.

Under a given turbulent velocity field defined by \cref{eq:linewidth_size}, there exists a family of quasi-equilibrium \ac{TES} solutions parametrized by $\xi_s$ and $p$.
\REV{These solutions are unstable when their maximum radial extent in terms of dimensionless $\xi$ exceeds the critical value $\xi_\mathrm{crit}$.}
\REV{For a given central density $\rho_c$, the critical radius in physical units is given by
\begin{equation}\label{eq:rcrit_rhoc}
    r_\mathrm{crit}(\rho_c) = \frac{\xi_\mathrm{crit}}{(4\pi)^{1/2}}\frac{c_s}{G^{1/2}\rho_c^{1/2}},
\end{equation}
where the corresponding mass within $r_\mathrm{crit}$ defines the critical mass
\begin{equation}
    M_\mathrm{crit}(\rho_c) = \frac{m_\mathrm{crit}}{(4\pi)^{1/2}}\frac{c_s^3}{G^{3/2}\rho_c^{1/2}}.
\end{equation}
Here, $\xi_\mathrm{crit}$ and $m_\mathrm{crit}$ are the dimensionless critical radius and critical mass, respectively, for which \citetalias{moon24} provides solutions as functions of $p$ and $\xi_s$} (see Figure 4 of \citetalias{moon24}; see also Equations (56)--(57) of \citetalias{moon24} for alternative expressions of $r_\mathrm{crit}$ and $M_\mathrm{crit}$ based on the edge and average densities).
\REV{Both $\xi_\mathrm{crit}$ and $m_\mathrm{crit}$ decrease with increasing $\xi_s$, approaching the well-known \ac{BE} limit $\xi_\mathrm{crit} = 6.45$ and $m_\mathrm{BE} = 15.7$ as $\xi_s\to\infty$.}

For $p=0.5$, \citetalias{moon24} found approximate relations to characterize how turbulence enhances the critical radius and mass for given mean density \REV{$\overline{\rho} \equiv 3M_\mathrm{crit}/(4\pi r_\mathrm{crit}^3) = 3\rho_c m_\mathrm{crit}/\xi_\mathrm{crit}^3$}:
\begin{equation}\label{eq:rcrit_fit}
    r_\mathrm{crit}(\overline{\rho}) \approx R_\mathrm{BE}(\overline{\rho}) \left( 1 + \frac{1}{2}\frac{\sigma_\mathrm{1D}^2}{c_s^2}\right)^{1/3},
\end{equation}
\begin{equation}\label{eq:mcrit_fit}
    M_\mathrm{crit}(\overline{\rho}) \approx M_\mathrm{BE}(\overline{\rho}) \left( 1 + \frac{1}{2}\frac{\sigma_\mathrm{1D}^2}{c_s^2}\right),
\end{equation}
which are valid within relative error of $5\%$ for $\sigma_\mathrm{1D}/c_s < 9.5$ and $\sigma_\mathrm{1D}/c_s < 13$, respectively, \REV{and where
\begin{equation}\label{eq:sigma1d}
  \sigma_\mathrm{1D} \equiv \frac{1}{\sqrt{3}}\left(\frac{\int \rho \vert\mathbf{v} - \mathbf{v}_\mathrm{com}\vert^2 d\mathcal{V}}{\int \rho\, d\mathcal{V}}\right)^{1/2}
\end{equation}
is the mass-weighted one-dimensional velocity dispersion relative to the center of mass velocity $\mathbf{v}_\mathrm{com} \equiv (\int \rho \mathbf{v}\,d\mathcal{V})/(\int \rho\,d\mathcal{V})$.
The volume integral is performed over a ball of radius $r_\mathrm{crit}$ in the context of the \ac{TES} model and our simulated cores.
While in general, ordered motions such as inflow, outflow, or rotation, can contribute to $\sigma_\mathrm{1D}$ in addition to random turbulent motions, such bulk components are assumed to be zero in the \ac{TES} model and found to be subdominant for our simulated cores except near the end of the collapse (see Appendix B of \citetalias{paperII}.)}
An important point to note is that \cref{eq:rcrit_fit,eq:mcrit_fit} do not simply substitute $c_s^2 \rightarrow c_s^2 + \sigma_\mathrm{1D}^2$ in \cref{eq:rbe,eq:mbe}, but rather, there are different functional dependences on turbulent and thermal velocities.\footnote{For the high-turbulence regime $\sigma_\mathrm{1D}/c_s\gtrsim 10$, a better fit replaces $0.5(\sigma_\mathrm{1D}/c_s)^2$ in both relations by $0.16(\sigma_\mathrm{1D}/c_s)^{2.5}$. However, this regime is not relevant to the critical cores found in our simulations.} 

We note that from the definitions in \cref{eq:linewidth_size} and \cref{eq:sigma1d}, the internal velocity dispersion, core radius, and sonic radius at the critical time are related by (see Appendix B of \citetalias{paperII} for details) 
\begin{equation}\label{eq:sigrcrs}
\frac{\sigma_\mathrm{1D}}{c_s} = \eta_d \left(\frac{3}{2p+3}\right)^{1/2}\left( \frac{r_\mathrm{crit}}{r_s}\right)^{p},    
\end{equation}
where $\eta_d$ is almost constant at $\eta_d \approx 0.9$, with very weak dependence on $p$ and $\sigma_\mathrm{1D}$.
Thus, trans-sonic cores with $p\sim 1/3-1/2$ have $r_\mathrm{crit}/r_s\sim 1$, \REV{and more generally $\sigma_\mathrm{1D}/c_s \approx 0.8(r_\mathrm{crit}/r_s)^p$}.

For a given power law index $p$, the \ac{TES} solutions show that critical cores have a dimensionless sonic radius $\xi_s$ (see \cref{eq:xi_def} for definition) that monotonically decreases with increasing $\sigma_\mathrm{1D}/c_s$ (see Figure 4(d) of \citetalias{moon24}).
For $p=0.5$, we find that an approximate relation \begin{equation}\label{eq:approx_xis_p05}
     \xi_s \approx 2.42 + 4\left(\frac{\sigma_\mathrm{1D}}{c_s}\right)^{-2},
 \end{equation}
 holds within a relative error of $1.85\%$ for $\sigma_\mathrm{1D} < 10 c_s$.

If a core with critical radius $r_\mathrm{crit}$ (based on its central density) and velocity dispersion $\sigma_\mathrm{1D}$ is in a state of turbulent quasi-equilibrium, with a linewidth--size index $p=0.5$, \cref{eq:rbe,eq:rcrit_fit} imply that its average density within $r_\mathrm{crit}$ would be approximately
\begin{equation}\label{eq:rhotes}
    \overline{\rho}_\mathrm{TES} \equiv 0.581\frac{c_s^2}{Gr_\mathrm{crit}^2}\left( 1 + \frac{1}{2}\frac{\sigma_\mathrm{1D}^2}{c_s^2}\right)^{2/3}.
\end{equation}
This can be regarded as a prediction of the \ac{TES} theory relating the radius, mean density, and velocity dispersion of critical cores.
As a generalization of the classical \ac{BE} relation, the prediction in \cref{eq:rhotes} can be compared to both simulations (as we shall do in this paper) and to observations.

\subsection{Hydrodynamic Calculations}

\REV{The initial condition of the turbulent cloud simulations conducted in \citetalias{paperII} consists of uniform density $\rho_0$ in a periodic box of side length $L_\mathrm{box}$,} with turbulent velocity perturbations randomly initialized from a Gaussian distribution with a $k^{-2}$ power spectrum (corresponding to a linewidth--size relation $\Delta v(l) \propto l^{1/2}$).
\REV{The velocity dispersion averaged over the simulation box, $\sigma_\mathrm{1D,box}$, is related to the specified three-dimensional Mach number by $\mathcal{M}_\mathrm{3D} = \sqrt{3}\sigma_\mathrm{1D,box}/c_s$.
}

\REV{In all our simulations, }we adopt an isothermal equation of state with 
\begin{equation}
    c_s = 0.188\,\mathrm{km\,s^{-1}} \left(\frac{T}{10\,\mathrm{K}}\right)^{1/2},
\end{equation}
assuming $\mu = 2.3 m_\mathrm{H}$ appropriate for $10\%$ helium abundance by number.
Apart from the dependence on a random seed \REV{for initializing the turbulent velocity field}, the initial conditions are fully specified with the two dimensionless parameters: the number of Jeans lengths initially contained in the domain, $L_\mathrm{box}/L_{J,0}$, and the root-mean-square Mach number $\mathcal{M}_\mathrm{3D}$ averaged within the cubic volume $L_\mathrm{box}^3$.
Here,
\begin{equation}\label{eq:ljeans}
  L_{J,0} \equiv \left(\frac{\pi c_s^2}{G\rho_0}\right)^{1/2} = 1.93\,\mathrm{pc}\left( \frac{T}{10\,\mathrm{K}} \right)^{1/2} \left( \frac{n_\mathrm{H,0}}{200\,\mathrm{cm}^{-3}} \right)^{-1/2}
\end{equation}
is the Jeans length at the cloud's average density $\rho_0 = 1.4 m_\mathrm{H} n_{\mathrm{H},0}$.
The corresponding mass and time scales are the Jeans mass
\begin{equation}\label{eq:mjeans}
  M_{J,0} \equiv \rho_0 L_{J,0}^3 = 49.9\,\mathrm{M}_\odot\,\left( \frac{T}{10\,\mathrm{K}} \right)^{3/2}\left( \frac{n_\mathrm{H,0}}{200\,\mathrm{cm}^{-3}} \right)^{-1/2} 
\end{equation}
and the Jeans time
\begin{equation}\label{eq:tjeans}
  t_{J,0} \equiv \frac{L_{J,0}}{c_s} = 10.0\,\mathrm{Myr}\,\left( \frac{n_\mathrm{H,0}}{200\,\mathrm{cm}^{-3}} \right)^{-1/2},
\end{equation}
which is related to the cloud's free-fall time
\begin{equation}\label{eq:tff0}
\begin{split}
  t_\mathrm{ff,0} &\equiv \left( \frac{3\pi}{32 G\rho_0} \right)^{1/2} = 0.306t_{J,0}\\
                  &= 3.08\,\mathrm{Myr} \left( \frac{n_\mathrm{H,0}}{200\,\mathrm{cm}^{-3}} \right)^{-1/2}.
\end{split}
\end{equation}

In \citetalias{paperII}, we presented results from 40 simulations conducted with $\mathcal{M}_\mathrm{3D} = 5$ (model \texttt{M5}) and 7 simulations with $\mathcal{M}_\mathrm{3D} = 10$ (model \texttt{M10}), varying the random seed for the initial cloud turbulence.
The box sizes for model \texttt{M5} and \texttt{M10} are set to $L_\mathrm{box} = 2L_{J,0}$ and $4L_{J,0}$, respectively, to fix the cloud-scale \REV{effective virial parameter at
\begin{equation}\label{eq:cloud_virial_parameter}
  \alpha_\mathrm{vir,box} \equiv \frac{5 \sigma_\mathrm{1D,box}^2 L_\mathrm{box}}{2G M_\mathrm{box}} = 1.66,
\end{equation}
where $M_\mathrm{box} = L_\mathrm{box}^3 \rho_0$ is the total mass in the computational domain.}
The chosen box size and Mach number set the average sonic radius,
\begin{equation}\label{eq:rs_cloud}
    r_{s,\mathrm{cloud}} \equiv \frac{9}{8}\frac{L_\mathrm{box}}{\mathcal{M}_\mathrm{3D}^2}
    = 0.579L_{J,0}\alpha_\mathrm{vir,box}^{-1/2}\mathcal{M}_\mathrm{3D}^{-1},
\end{equation}
at which turbulent velocity dispersion following the cloud's mean linewidth--size relation equals the sound speed; the factor $9/8$ accounts for the conversion between three-dimensional volume-average $\mathcal{M}_\mathrm{3D}$ and one-dimensional shell-average $\left<\delta v_r^2\right>_\rho^{1/2}$ (see Equations \eqsigmaoneddef and \eqlambdastors in \citetalias{paperII}).

\REV{
We choose physical and numerical parameters for our simulations by requiring both the overall sonic scale of the turbulence and the size of self-gravitating cores that subsequently form to be resolved.}
Cores that develop in simulations of turbulent clouds have densities comparable to the post-shock value at the global Mach number of the cloud, $\rho \sim \rho_\mathrm{ps} = \mathcal{M}_\mathrm{3D}^2\rho_0$. 
\cref{eq:rs_cloud} thus indicates that, for realistic cloud models with $\alpha_\mathrm{vir,box} \gtrsim 2$, resolving the sonic radius in ambient gas poses a similar challenge to resolving dense cores, which have typical sizes on the order of $R_\mathrm{BE}(\rho_\mathrm{ps}) = 0.430L_{J,0}\mathcal{M}_\mathrm{3D}^{-1}$.
With the goal of having $r_{s,\mathrm{cloud}}$ sufficiently resolved \REV{everywhere} and also having cores resolved down to a mass $M_\mathrm{BE}(2\rho_\mathrm{ps})$ we \REV{adopt uniform resolution with} the number of cells per dimension $N = 1024$ for model \texttt{M10} and $N = 512$ for \texttt{M5} (see Section \secresolution of \citetalias{paperII} for details on the resolution requirements).
The resulting cell size normalized to the average Jeans length is $\Delta x/L_{J,0} = (L_\mathrm{box}/N)/L_{J,0} = 3.91\times 10^{-3}$ for both models.

\REV{Since we adopt fixed spatial resolution, runaway gravitational collapse cannot be followed to later stages where the gravitational length scales fall below the grid scale.
To mitigate this difficulty, we adopt a sink particle algorithm in which each collapsing center is replaced by an accreting particle.
Numerical details of our sink particle algorithm are described in \citetalias{paperII}.
We stop each simulation when the total mass in sink particles exceeds  $0.15M_\mathrm{box}$.
}

\REV{
We emphasize that the calculations carried out in \citetalias{paperII} and analyzed here are designed as a set of numerical experiments to test the theoretical ideas outlined in \citetalias{moon24}.
Additional physics, such as magnetic fields and protostellar outflows, can be included in future numerical models to make them more realistic, and therefore more directly comparable to observations. For present purposes, our intention is to keep the simplest possible setup, to facilitate physical understanding and to establish a baseline for future developments.}

\subsection{Definition of Critical Cores}

\REV{
We begin our analysis by identifying the centroids of all cores with local gravitational potential minima, which are then linked through time based on position and velocity (see Section \seccoretracking of \citetalias{paperII} for details).
For each core center, we construct radial profiles of various physical quantities using \cref{eq:mass-weighted-angle-average,eq:volume-weighted-angle-average}, and monitor their temporal evolution.
Because instability is expected to set in at $r_\mathrm{crit}$ according to the \ac{TES} model, we define the critical time $t_\mathrm{crit}$ by}
\begin{equation}\label{eq:tcritdef}
    t_\mathrm{crit} \equiv \min\left\{t^* \mid F_\mathrm{net}(r_\mathrm{crit}) < 0\;\forall t\in (t^*, t_\mathrm{coll})\right\}.
\end{equation}
That is, the net radial force $F_\mathrm{net}(r_\mathrm{crit})$ acting on a ball of radius $r_\mathrm{crit}$ becomes negative at $t_\mathrm{crit}$ and stays negative until the end of the collapse at $t_\mathrm{coll}$ (see Equation \eqlageom of \citetalias{paperII} for \REV{the precise definition of $F_\mathrm{net}$ and its relation to the radial acceleration.)
Figure \figcoreevolution of \citetalias{paperII} shows that $t_\mathrm{crit}$ corresponds to the epoch when the average infall speed starts to accelerate, signaling the onset of runaway collapse.
}

\REV{
To calculate $r_\mathrm{crit}$ for simulated cores, we first fit \cref{eq:linewidth_size} to the actual profile of $\left<\delta v_r^2\right>_\rho^{1/2}/c_s$ for each core to determine $\xi_s$ and $p$, where we use the measured central density $\rho_c$ to convert the physical $r$ to dimensionless $\xi$ using \cref{eq:xi_def}.
We then calculate the dimensionless critical radius $\xi_\mathrm{crit}$ by solving the \ac{TES} equations\footnote{We provide our Python implementation in \url{https://github.com/sanghyukmoon/tesphere}.} and convert to $r_\mathrm{crit}$ by using \cref{eq:rcrit_rhoc}.
We also calculate the critical mass $M_\mathrm{crit}$ for each core in an analogous manner.
\citetalias{paperII} found that, at $t_\mathrm{crit}$, the critical radius becomes comparable to the ``tidal radius'' defined in terms of the local gravitational potential geography.}

For the collapse time of each core, we use the following working definition:
\begin{equation}\label{eq:tcolldef}
    t_\mathrm{coll} \equiv \text{The time of each sink particle formation}.
\end{equation}
The duration of collapse for each core is then defined by
\begin{equation}\label{eq:dtcoll}
  \Delta t_\mathrm{coll}\equiv t_\mathrm{coll} - t_\mathrm{crit},
\end{equation}
which can be measured for individual cores.
Note that \cref{eq:tcritdef,eq:tcolldef} provide quantitative definition of the critical time and the collapse time described in \cref{sec:intro}.

We consider a self-gravitating core with radius $r_\mathrm{crit}$ resolved if
\begin{equation}\label{eq:resolvedness_criterion}
    r_\mathrm{crit} > N_\mathrm{core,res}\Delta x,
\end{equation}
where our standard choice is $N_\mathrm{core,res} = 8$ unless otherwise mentioned.
We note that this is a conservative choice amounting to resolving a core volume with $4\pi N_\mathrm{core}^3/3 = 2145$ resolution elements.
A core would still be marginally resolved with less restrictive choice of $N_\mathrm{core,res} = 4$.
\REV{Below this limit, turbulence within cores would be significantly affected by numerical dissipation and their evolution is not reliable.
For example, had the resolution been higher and the dissipation lower, a core could have been dispersed by turbulence instead of initiating collapse.
With our default choice of $N_\mathrm{core,res} = 8$, there are $83$ and $79$ cores in our entire ensemble of simulations for model \texttt{M5} and \texttt{M10}, respectively.
These numbers increase to $107$ and $129$ when $N_\mathrm{core,res} = 4$ is adopted.
It is expected that the number of resolved cores per simulation would increase with increasing numerical resolution, until the entire \ac{CMF} is resolved.
In \cref{sec:CMF}, we will present evidence that the peak of the \ac{CMF} is marginally resolved in model \texttt{M5}.
We will further discuss numerical requirements for resolving the \ac{CMF} in \autoref{app:appendix}.
}

Given that the radius of the smallest resolvable core would be $R_\mathrm{BE} = N_\mathrm{core,res}\Delta x$, one can use the corresponding \ac{BE} mass to derive the minimum resolvable mass:
\begin{equation}\label{eq:Mmin_given_dx}
    M_\mathrm{min} = 2.43 \frac{c_s^2}{G} N_\mathrm{core,res}\Delta x.
\end{equation}
For simulations adopting a fixed mass resolution $\Delta m$, the equivalent minimum resolvable mass may be written as
\begin{equation}\label{eq:Mmin_given_dm}
    M_\mathrm{min} = \frac{4\pi N_\mathrm{core,res}^3}{3} \Delta m.
\end{equation}

\REV{
In the remaining text, we use the term ``critical core'' to denote a ball of radius $r_\mathrm{crit}$ at $t = t_\mathrm{crit}$.
The radius and mass of a critical core is defined by
\begin{equation}\label{eq:Rcoredef}
  R_\mathrm{core} \equiv r_\mathrm{crit}(t=t_\mathrm{crit}),
\end{equation}
\begin{equation}\label{eq:Mcoredef}
  M_\mathrm{core} \equiv \iiint_{r < R_\mathrm{core}} \rho(t=t_\mathrm{crit})\,dV,
\end{equation}
where all quantities are evaluated at $t = t_\mathrm{crit}$ for each core.
}

\section{Properties of Critical Cores}\label{sec:critical_core_properties}

In this section, we present the physical properties of \REV{critical} cores in models \texttt{M5} and \texttt{M10}.
\REV{We only include resolved cores with $R_\mathrm{core} > 8\Delta x$ unless otherwise stated.}

\subsection{Radial Density Profiles}\label{sec:radprof}

\begin{figure*}[htpb]
  \epsscale{1.18}
  \plotone{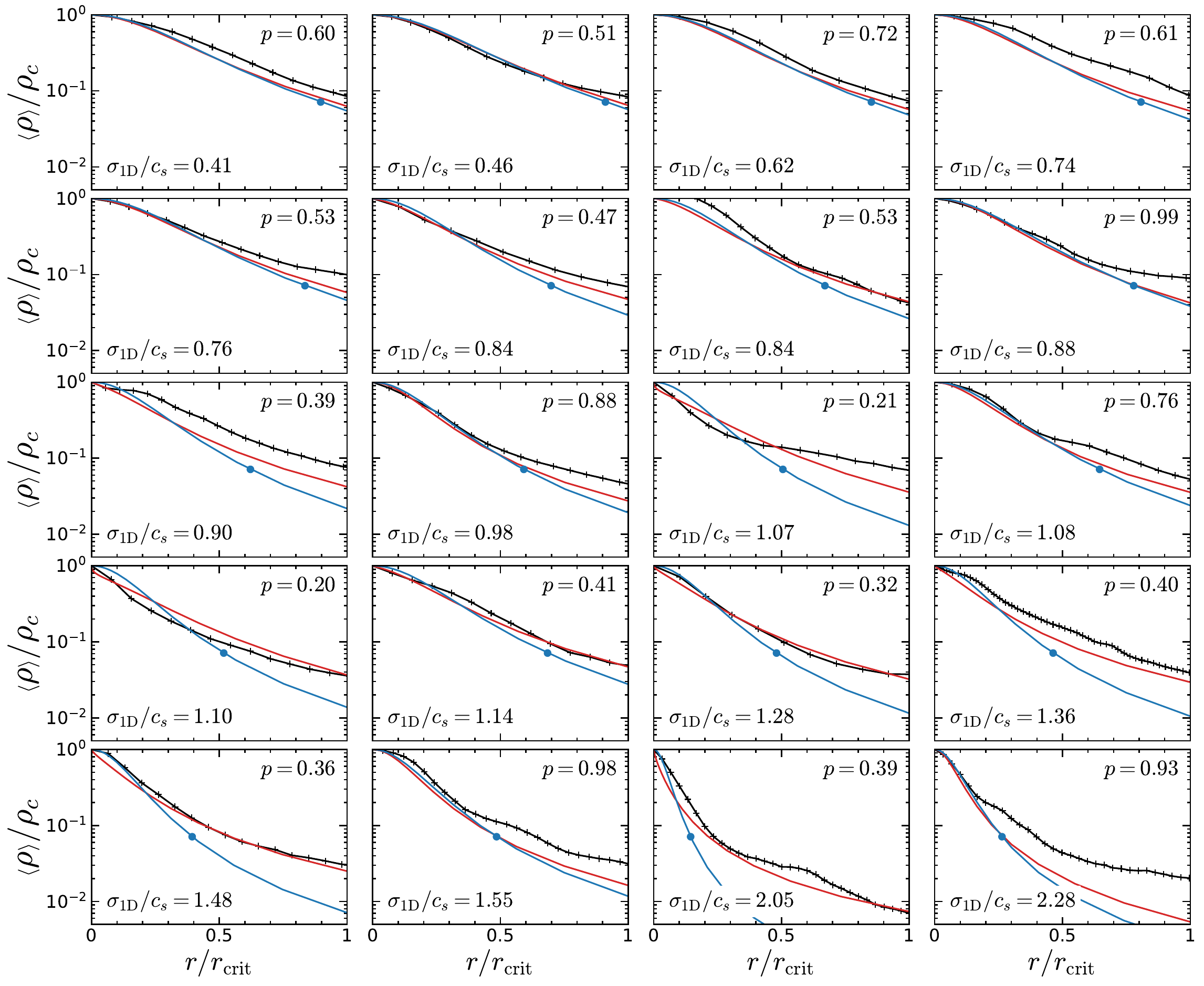}
  \caption{Radial density profiles of 20 randomly selected critical cores in model \texttt{M10}, arranged in an increasing order of $\sigma_\mathrm{1D}$ from left to right and top to bottom.
  In each panel, the black line with markers plots the measured density profile while the red line shows the analytic equilibrium solution of the critical \ac{TES} constructed from the measured $\rho_c$, $r_s$, and $p$ (i.e., the red line is not a fit to the black line).
  For comparison, we also plot the similar profile of the \ac{BE} sphere in the blue line using the measured $\rho_c$ but assuming only thermal pressure (i.e.~$r_s\to \infty$).
  The blue circle marks the location of $R_\mathrm{BE}$.
  In each panel, the density and radius are normalized by $\rho_c$ and $R_\mathrm{core} = r_\mathrm{crit}$, respectively.
  The mass-weighted average velocity dispersion $\sigma_\mathrm{1D}$ and the power-law index of the fitted linewidth--size relation $p$ are annotated in each panel.}
  \label{fig:radial_profiles_tcrit}
\end{figure*}

\cref{fig:radial_profiles_tcrit} plots the normalized radial density profiles measured at $t = t_\mathrm{crit}$, for 20 randomly selected cores in model \texttt{M10}.
For comparison, we also plot the theoretical \ac{TES} and \ac{BE} solutions obtained using the measured central density $\rho_c$, sonic radius $r_s$, and power law index $p$ (for the \ac{BE} sphere, we assume $r_s\to \infty$).
The latter two quantities are obtained by fitting \cref{eq:linewidth_size} to the measured profile of $\left<\delta v_r^2\right>_\rho^{1/2}$, within the tidal radius defined by the distance to the nearest saddle point of the gravitational potential (see \citetalias{paperII}).
\cref{fig:radial_profiles_tcrit} indicates that while the measured profiles are broadly consistent with the \ac{TES} solutions, they cannot be described by \ac{BE} spheres unless $\sigma_\mathrm{1D}$ (see \cref{eq:sigma1d} for definition) is small compared to $c_s$ or the linewidth--size index $p$ (see \cref{eq:linewidth_size}) is large.
For the cases with supersonic turbulence within a core, density decreases at large radii much less in the simulated cores (and in the \ac{TES} solutions) than in the \ac{BE} solutions that match the profiles at small radius.
Some cores do show moderate discrepancies from \ac{TES} profiles for one or a combination of the following reasons: Departure from the spherical symmetry or the presence of nearby structures that make the gravitational force deviate from $GM/r^2$; traveling shock waves that intermittently throw a core out of equilibrium; strong converging flows building a core in less than a sound crossing time (we come back to this in \cref{fig:building_time}); non-negligible rotation; a linewidth--size relation which does not conform to a single power law.
Nonetheless, the overall agreement between radial density profiles of the simulated cores and \ac{TES} solutions suggests that quasi-steady equilibrium structures supported by thermal and turbulent pressures naturally emerge from the interplay between supersonic turbulence and self-gravity.

\subsection{Local Linewidth--size Relations and Correlation with Density}\label{sec:linewidth_size}

\begin{figure}[htpb]
  \epsscale{1.18}
  \plotone{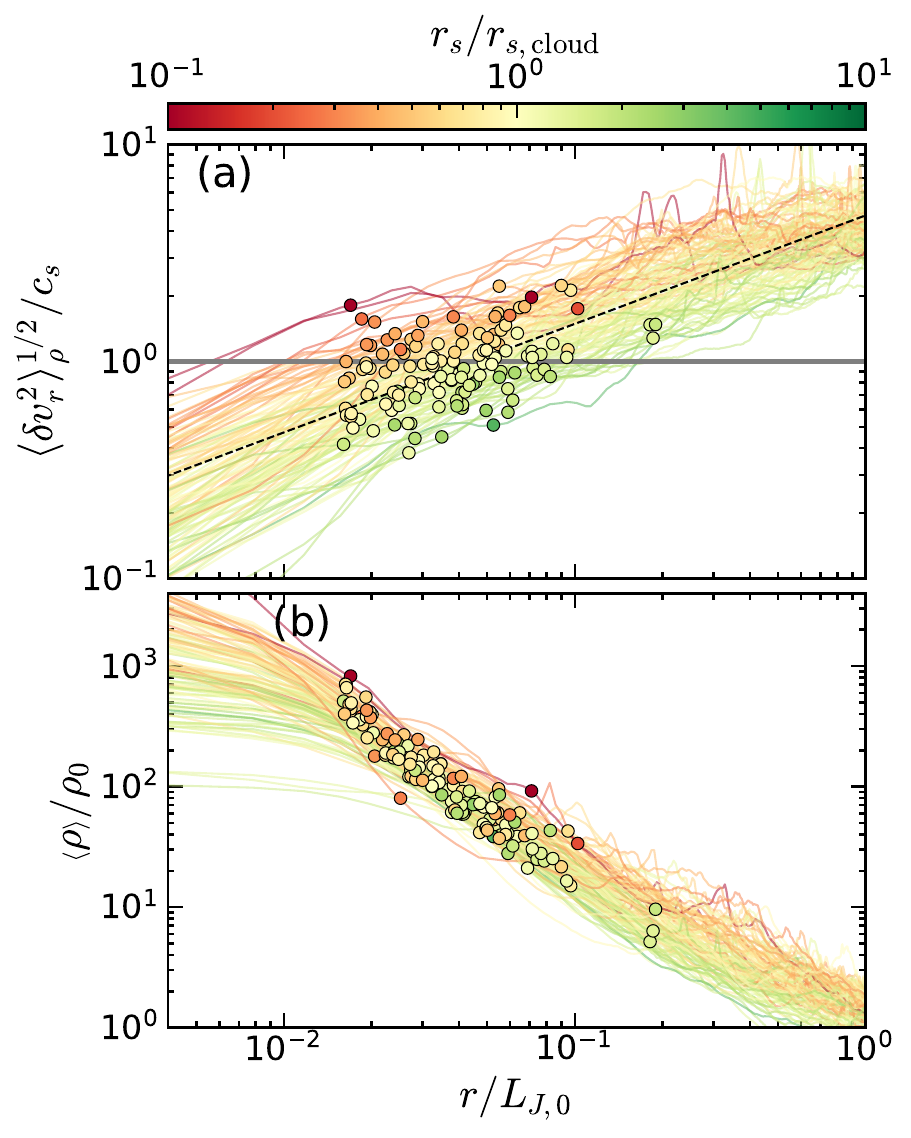}
  \caption{Radial profiles of (a) turbulent velocity dispersion and (b) density, for the critical cores in model \texttt{M10}.
  Individual lines correspond to the profiles constructed starting from the center of each core, where we include the cores with $r_\mathrm{crit} \ge 4\Delta x$ (i.e., $N_\mathrm{core,res} = 4$).
  The critical radius $r_\mathrm{crit}$ is marked with circles.
  The black dashed line in panel (a) plots the average linewidth--size relation, $\left< \delta v_r^2\right>_\rho^{1/2}/c_s = (r / r_{s,\mathrm{cloud}})^{0.5}$, where the colors represent the ratio of the local sonic radius $r_s$ to the average $r_{s,\mathrm{cloud}}$ (i.e., the local deviations from the mean turbulent scaling relation).
  The intersection points between each colored line and the horizontal gray line representing $\left<\delta v_r^2\right>_\rho^{1/2}/c_s = 1$ correspond to the local sonic radius $r_s$ for each core.}
  \label{fig:density_velocity_profiles}
\end{figure}

\begin{figure}[htpb]
  \epsscale{1.18}
  \plotone{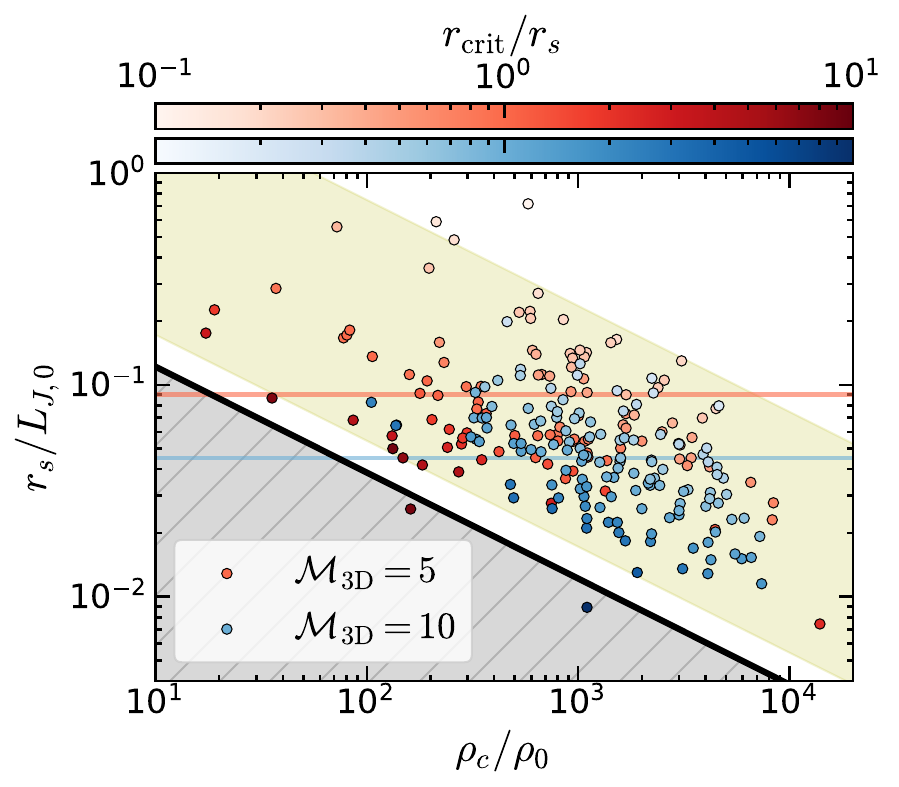}
  \caption{Local sonic radius versus central density measured for selected individual cores from model \texttt{M5} (red) and \texttt{M10} (blue), satisfying the resolution criterion $N_\mathrm{core,res} = 4$.
  Black line plots \cref{eq:rhocmin_local}, such that a \ac{TES} with $p=0.5$ is stable at all radii when $r_s$ and $\rho_c$ falls in the gray shaded region.
  Horizontal red and blue lines correspond to $r_{s,\mathrm{cloud}}$ (defined in \cref{eq:rs_cloud}) for model \texttt{M5} and \texttt{M10}, respectively.
  Circles are colored by $r_\mathrm{crit}/r_s$, where  $\sigma_\mathrm{1D}/c_s \approx 0.8(r_\mathrm{crit}/r_s)^{1/2}$ from \cref{eq:sigrcrs}.
  The yellow shaded band is the region where $\sigma_\mathrm{1D}/c_s$ is between $0.3$ (upper edge) and $2$ (lower edge).}
  \label{fig:rs_rhoc_correlation}
\end{figure}

The comparison in \cref{fig:radial_profiles_tcrit} suggests that turbulence affects density structure and stability to varying extents across different cores.
In this section, we demonstrate how the local variations of the turbulent scaling relation affect the critical density at which collapse is triggered.
To illustrate the variations in the locally-constructed linewidth--size relations, in \cref{fig:density_velocity_profiles}(a) we plot the turbulent velocity dispersion $\left<\delta v_r^2\right>_\rho^{1/2}$ measured for each critical core as a function of growing radius from the core center (see \cref{eq:linewidth_size}).
The colors show the ratio $r_s/r_{s,\mathrm{cloud}}$, i.e., the local sonic radius relative to the cloud average defined in \cref{eq:rs_cloud}.
\cref{fig:density_velocity_profiles}(a) shows that the turbulent velocity dispersion measured at a given size scale varies up to a factor of a few above and below the average relation given by $c_s(r/r_{s,\mathrm{cloud}})^{0.5}$.
Alternatively, it implies that the locally measured sonic radius, $r_s$, exhibits order of magnitude variations around $r_{s,\mathrm{cloud}}$.

Large spatial variations in the strength of turbulence shown in \cref{fig:density_velocity_profiles}(a) suggest that the necessary density for cores to collapse would differ considerably from core to core.
\cref{fig:density_velocity_profiles}(b) plots the density profiles of critical cores normalized to the average cloud density $\rho_0$ and the associated Jeans length $L_{J,0}$, with the same color coding as in \cref{fig:density_velocity_profiles}(a).
The noticeable vertical color gradients suggest that cores forming in more turbulent regions (i.e., smaller $r_s$) overall have higher densities when compared at a given radius.

To illustrate this point more clearly, in \cref{fig:rs_rhoc_correlation} we plot the central density $\rho_c$ versus the local sonic radius $r_s$ for all critical cores in models \texttt{M5} and \texttt{M10}.
It shows that, for critical cores, $r_s$ and $\rho_c$ are anticorrelated, meaning that cores in regions of strong turbulence need to reach higher densities to initiate collapse.
The distribution shown in \cref{fig:rs_rhoc_correlation} also exhibits scatter in the direction perpendicular to the overall anticorrelation, which, together with the overall anticorrelation, can be understood in the framework of \acp{TES} as follows.

Assuming $p=0.5$, \cref{eq:xi_def,eq:approx_xis_p05} yield an expression for the central density required for a core to collapse, depending on its level of turbulence and the local sonic radius:
\begin{equation}\label{eq:rhocrit}
    \frac{\rho_c}{\rho_0} \approx \frac{1}{4\pi^2}\left(\frac{r_s}{L_{J,0}}\right)^{-2} \left[2.42 + 4\left(\frac{\sigma_\mathrm{1D}}{c_s}\right)^{-2}\right]^2.
\end{equation}
We note that the term in the square bracket can be alternatively expressed by $2.42 + 6.6r_s/r_\mathrm{crit}$, using \cref{eq:sigrcrs} with $\eta_d = 0.9$ and $p=0.5$.
\REV{Additionally, it is worth noting that the mean cloud density $\rho_0$ only enters \cref{eq:rhocrit} because we write $r_s$ in units of $L_{J,0} = c_s (\pi/G\rho_0)^{1/2}$.}
Thus, \cref{eq:rhocrit} indicates that $\rho_c$ is overall proportional to $r_s^{-2}$, with scatter from varying $\sigma_\mathrm{1D}/c_s$ modulated by relative core size $r_\mathrm{crit}/r_s$, consistent with the trend observed in \cref{fig:rs_rhoc_correlation}.
At fixed $r_s$, larger and more turbulent cores (which go together following  $\sigma_\mathrm{1D}/c_s\approx 0.8 (r_\mathrm{crit}/r_s)^p$) have lower $\rho_c/\rho_0$. As emphasized in \citetalias{moon24} and \citetalias{paperII}, core size is subject to the local gravitational potential environment.

It is worth noting that almost all critical cores fall under the region of parameter space where $\sigma_\mathrm{1D}/c_s \in [0.3, 2]$ (shown as yellow shaded band in \cref{fig:rs_rhoc_correlation}).
That is, we find neither purely thermal cores with $\sigma_\mathrm{1D} \ll c_s$, nor highly turbulent cores with $\sigma_\mathrm{1D} \gg c_s$.
The center of the distribution shown in \cref{fig:rs_rhoc_correlation} lies roughly along $\xi_s \sim 9$, corresponding to $\sigma_\mathrm{1D}/c_s = 0.78$.
For a $p=0.5$ \ac{TES}, the critical radius becomes identical to the sonic radius when $\xi_s \sim 9$ (see Figure 4(a) of \citetalias{moon24}).
This means that cores in our simulations on average have $r_\mathrm{crit} \sim r_s$ at the onset of collapse.
We discuss physical reasons for this below.

\citetalias{paperII} found that collapse occurs when $r_\mathrm{crit}$ decreases below the tidal radius $r_\mathrm{tidal}$, which is set by local geography of the gravitational potential.
Our results therefore imply that critical cores generally satisfy $r_\mathrm{crit} \sim r_s \sim r_\mathrm{tidal}$.
Indeed, $r_s$ is the scale at which the nonlinear structures are expected, which may in turn determine the size of the local potential well characterized by $r_\mathrm{tidal}$.

We additionally note that, for a given $r_s$, there exists a minimum central density below which all \ac{TES} solutions are stable \citepalias{moon24}.
For $p = 0.5$, this minimum can be obtained by taking the $\sigma_\mathrm{1D} \to \infty$ limit in \cref{eq:rhocrit}, leading to
\begin{equation}\label{eq:rhocmin_local}
    \frac{\rho_{c,\mathrm{min}}}{\rho_0} = 0.148 \left(\frac{r_s}{L_{J,0}}\right)^{-2},
\end{equation}
\REV{or equivalently $\rho_{c,\mathrm{min}}=0.46 (c_s/r_s)^2/G$.}
This is plotted in a solid black line in \cref{fig:rs_rhoc_correlation}.
In the presence of turbulence satisfying \cref{eq:linewidth_size} with $p=0.5$, the theory predicts that collapse cannot occur in the shaded region below this borderline.
The absence of critical cores from our simulations in the shaded region is consistent with this theoretical prediction.
As shown in Figure \figrsrhocevolution of \citetalias{paperII}, cores generally form in this stable regime and then evolve diagonally through the borderline, until the critical conditions are satisfied at various locations within the yellow shaded region of the parameter space shown in \cref{fig:rs_rhoc_correlation}.

\begin{figure*}[htpb]
    \epsscale{1.18}
    \plotone{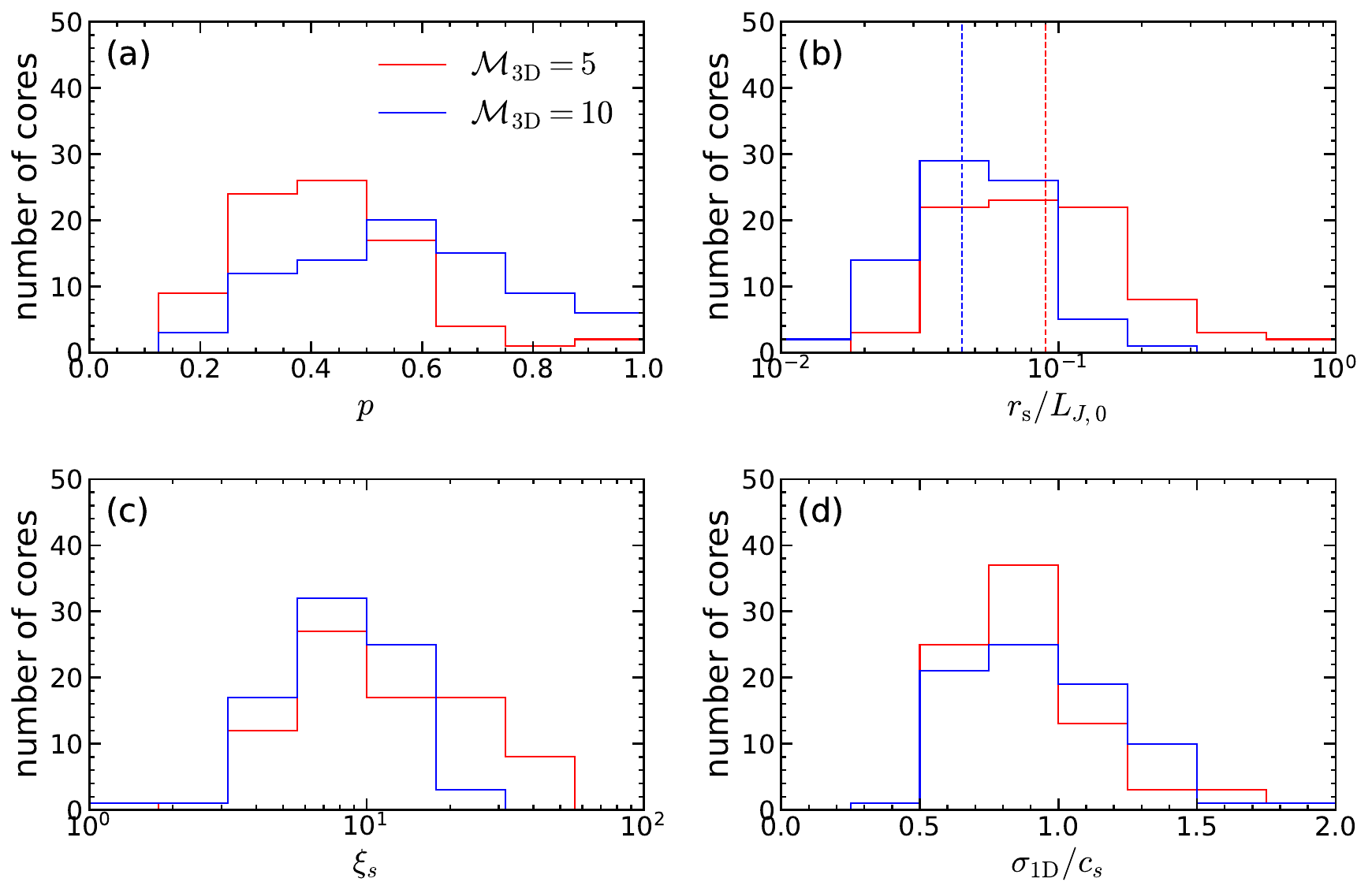}
    \caption{Distributions of dynamical properties in all critical cores from models \texttt{M5} (red) and \texttt{M10} (blue).
    Shown are (a) the power-law exponent $p$ of the linewidth--size relation, (b) the sonic radius $r_s$, (c) the dimensionless sonic radius $\xi_s$  (see \cref{eq:xi_def}), and (d) the mass-weighted average velocity dispersion $\sigma_\mathrm{1D}$ (\cref{eq:sigma1d}).
    The vertical dotted lines in panel (b) mark the initial sonic radius $r_{s,\mathrm{cloud}}$ defined in (\cref{eq:rs_cloud})
    which is expected from the large-scale $k^{-2}$ velocity power spectrum.}
  \label{fig:tes_parameters}
\end{figure*}

\subsection{Distribution of Turbulence Parameters and Core Densities}\label{sec:core_properties}

\cref{fig:tes_parameters} plots the distribution of $p$, $r_s$, $\xi_s$, and $\sigma_\mathrm{1D}$ from all critical cores in models \texttt{M5} and \texttt{M10}.
The median values of $p$ (the index of the fitted linewidth--size relation in \cref{eq:linewidth_size}) are $0.41$ and $0.56$ for model \texttt{M5} and \texttt{M10}, respectively, roughly consistent with the expected average slope of $0.5$ resulting from the initial velocity power spectrum.
However, the distribution is rather broad, such that individual cores can have diverse values of $p$ ranging between $\sim 0.2\text{--}0.8$.
Cores in model \texttt{M5} have overall slightly smaller values of $p$ compared to those forming in model \texttt{M10}.

Due to higher initial $\mathcal{M}_\mathrm{3D}$,
the sonic radius of the cores is generally smaller in model \texttt{M10}.
The median values are $r_s = 8.7 \times 10^{-2} L_{J,0}$ and $5.0 \times 10^{-2} L_{J,0}$ for model \texttt{M5} and \texttt{M10}, respectively, entirely consistent with what expected from the initial box-scale Mach number (\cref{eq:rs_cloud}).\footnote{Although the box-averaged Mach number has decayed from its initial $\mathcal{M}_\mathrm{3D}$ by the time most cores form and collapse, we find the energy loss is predominantly at large scales and the velocity structure function at core scales is in fact consistent with the initial condition (see \cref{fig:density_velocity_profiles}(a) as well as Figure \figlinewidthsize of \citetalias{paperII}).}

While cores in model \texttt{M10} have overall smaller $r_s$ than model \texttt{M5}, they tend to have higher $\rho_c$, presumably due to the stronger turbulent compression of core-forming regions at higher cloud-scale Mach number.
This makes the distribution of the dimensionless sonic radius $\xi_s = (4\pi G \rho_c)^{1/2} r_s/c_s$ (which determines the structure of forming cores; see \citetalias{moon24}) not very different between models \texttt{M5} and \texttt{M10}.
The median and $\pm 34.1$th percentile range of $\xi_s$ for the entire ensemble of cores is $9.5^{+9.2}_{-4.3}$.

\begin{figure*}[htpb]
  \epsscale{1.18}
  \plotone{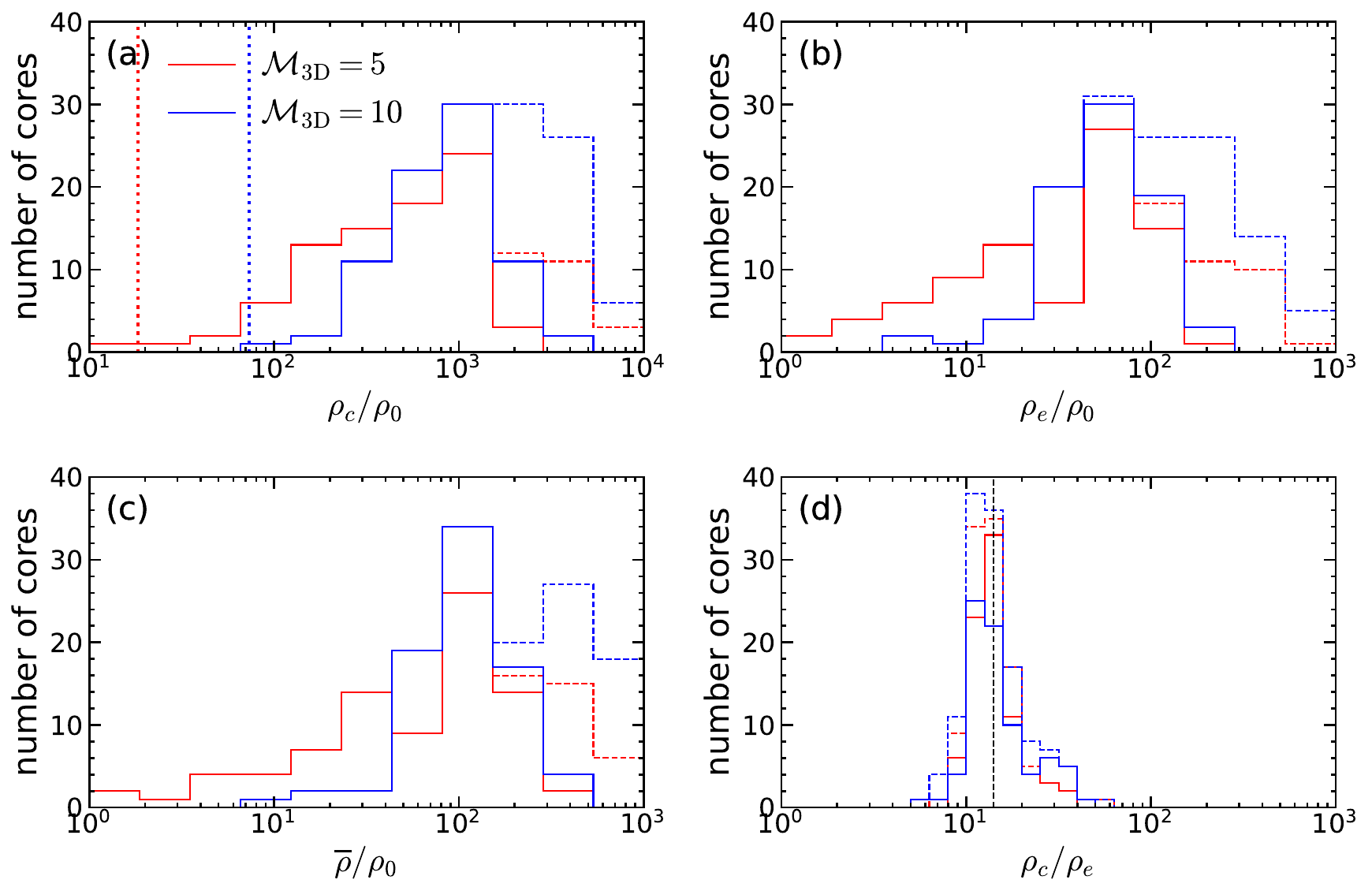}
  \caption{Distribution of density values in all critical cores from models \texttt{M5} (red) and \texttt{M10} (blue), satisfying the resolution criterion (\cref{eq:resolvedness_criterion}) with the standard choice of $N_\mathrm{core,res} = 8$ (solid histograms) and the relaxed choice $N_\mathrm{core,res} = 4$ (dashed histograms).
  Shown are (a) center, (b) edge, (c) mean densities relative to the mean density in the box, and (d) the center-to-edge density contrast.
  In (a), vertical dotted lines plot the theoretical prediction for the minimum central density allowing instability, assuming $p=0.5$ (\cref{eq:rhocmin}).
  The vertical dotted line in the panel (d) marks $\rho_c/\rho_e = (\rho_c/\rho_e)_\mathrm{BE} = 14$.}
  \label{fig:core_densities}
\end{figure*}

\cref{fig:tes_parameters}(d) shows that most cores in models \texttt{M5} and \texttt{M10} have subsonic or transonic velocity dispersions, as noted in connection to \cref{fig:rs_rhoc_correlation}.
The level of turbulence in our critical cores is consistent with the kinematics of observed prestellar cores \citep[e.g.,][]{cwlee99,cwlee01,foster09,friesen09,friesen10,lee14,storm16,keown17,kirk17,tang18,chen19,chung19,kerr19,choudhury21,chung21,li23,yoo23}.
The median and $\pm 34.1$th percentile range of $\sigma_\mathrm{1D}/c_s$ is $0.88^{+0.30}_{-0.18}$.

\cref{fig:core_densities} plots the distribution of densities of cores at their critical time: Central $\rho_c$, edge $\rho_e$ (defined as the value at $r=R_\mathrm{core}$), and mean $\overline{\rho} = M_\mathrm{core} / (4\pi R_\mathrm{core}^3 / 3)$, as well as the center-to-edge contrast $\rho_c/\rho_e$.
This makes clear that cores form at a wide range of densities, with a typical value around $\overline{\rho} \sim 10^2\rho_0$, corresponding to $\overline{n}_\mathrm{H_2} \sim 10^4\,\mathrm{cm}^{-3}$ for a cloud with average molecular hydrogen number density $n_\mathrm{H_2,0} = 10^2\,\mathrm{cm}^{-3}$.
A caveat is that the distribution is likely to become incomplete somewhere in the regime $\overline{\rho}/\rho_0 \gtrsim 10^2$--$10^3$; we refer the reader to Section \secresolution of \citetalias{paperII} for a discussion of our resolution criteria.
The distribution of center-to-edge density contrast is narrow, peaking at $\rho_c/\rho_e\sim 10\text{--}30$.
Some cores have a center-to-edge contrast of less than $14$ because they have excess mass at their edge compared to isolated \ac{TES} solutions (e.g., \cref{fig:radial_profiles_tcrit}).

It is interesting to reexamine the distribution of $\rho_c$ in models \texttt{M5} and \texttt{M10} by using \cref{eq:rhocrit}.
Neglecting the local variations of $r_s$ and instead adopting the ``average'' value $r_s = r_{s,\mathrm{cloud}}$ defined in \cref{eq:rs_cloud}, one can relate the required central density  to a cloud's virial parameter and Mach number as
\begin{equation}\label{eq:rhocrit_cloud}
    \frac{\rho_c}{\rho_0} \approx 7.56\times 10^{-2}\left[2.42 + 4\left(\frac{\sigma_\mathrm{1D}}{c_s}\right)^{-2}\right]^2 \alpha_\mathrm{vir,box}\mathcal{M}_\mathrm{3D}^2,
\end{equation}
where an implicit assumption is $p=0.5$ inherited from using \cref{eq:approx_xis_p05}.
For a cloud with given $\alpha_\mathrm{vir,box}$ and $\mathcal{M}_\mathrm{3D}$, \cref{eq:rhocrit_cloud} has a lower bound,
\begin{equation}\label{eq:rhocmin}
    \frac{\rho_{c,\mathrm{min}}}{\rho_0} = 0.443\alpha_\mathrm{vir,box}\mathcal{M}_\mathrm{3D}^2,
\end{equation}
in the limit of $\sigma_\mathrm{1D} \to \infty$.
For models \texttt{M5} and \texttt{M10}, \cref{eq:rhocmin} yields $\rho_{c,\mathrm{min}} = 18\rho_0$ and $74\rho_0$, respectively, which are intriguingly similar to the lower edges of the central density distribution shown in \cref{fig:core_densities}(a).
For transonic cores with \REV{$r_\mathrm{crit} = r_s$ (i.e.,  $\sigma_\mathrm{1D} \approx 0.8c_s$, }representative in \cref{fig:tes_parameters}(d)), \cref{eq:rhocrit_cloud} yields $\rho_c = 236\rho_0$ and $943\rho_0$ for models \texttt{M5} and \texttt{M10}, respectively, which are closer to typical central densities of cores.

\subsection{Core Mass and Radius Distributions}\label{sec:CMF}

\begin{figure*}[htpb]
  \epsscale{1.18}
  \plotone{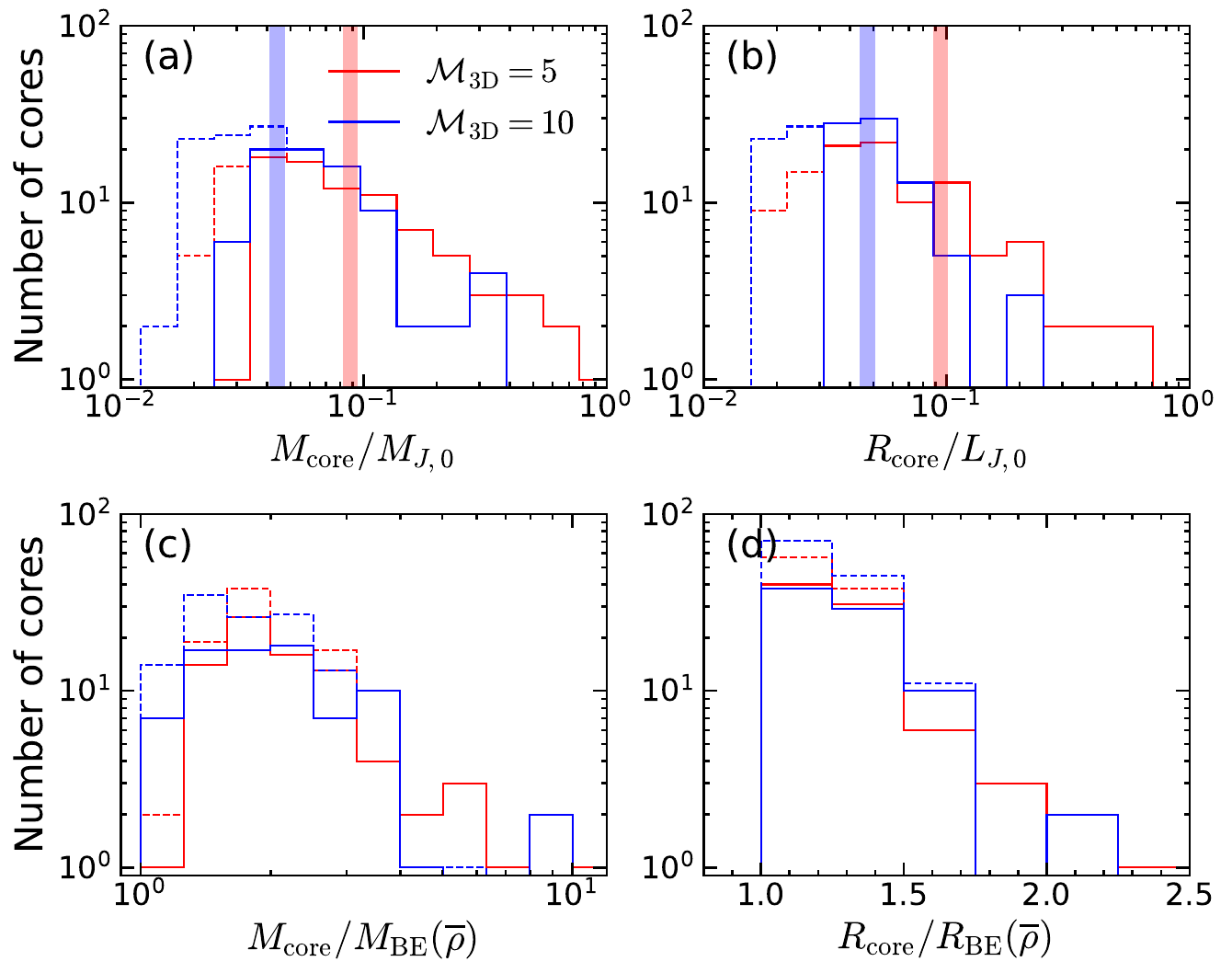}
  \caption{Distribution of mass and radius of critical cores in models \texttt{M5} (red) and \texttt{M10} (blue), satisfying the resolution criterion (\cref{eq:resolvedness_criterion}) with the standard choice of $N_\mathrm{core,res} = 8$ (solid histograms) and the relaxed choice $N_\mathrm{core,res} = 4$ (dashed histograms).
  (a) The \ac{CCMF}; bin lower limits for mass are at $M_\mathrm{min}$ from \cref{eq:Mmin_given_dx} for a chosen $N_\mathrm{core,res}$.
  (b) The distribution of core radius, where by definition $R_\mathrm{core} = r_\mathrm{crit}$ at $t = t_\mathrm{crit}$.
  Lower limits are at $N_\mathrm{core,res}\Delta x$.
  For (a) and (b)  shaded bands mark  $M_\mathrm{char,trb}$ and $R_\mathrm{char,trb}$ defined in \cref{eq:mcrit_ps} and \cref{eq:rcrit_ps} for $\mathcal{M}_\mathrm{3D} = 5$ and $10$, respectively  (see text for details). 
  (c) The distribution of the ratio between the core mass and the \ac{BE} mass evaluated at the average core density.
  (d) The distribution of the ratio between the core radius and the \ac{BE} radius evaluated at the average core density.
  }
  \label{fig:mass_function}
\end{figure*}

\cref{fig:mass_function}(a) plots the mass function of critical cores, defined as the number of cores per logarithmic interval of $M_\mathrm{core}$, for models \texttt{M5} and \texttt{M10}.
We introduce a new term \ac{CCMF} to distinguish our mass function, which is specifically constructed at $t_\mathrm{crit}$ of each core, from observed \acp{CMF} consisting of cores at various evolutionary stages.
We also plot the distribution of the core radius in \cref{fig:mass_function}(b).
We remind the reader that these cores are defined at their own critical time, which occurs at a different point in the simulation for each core.
Although the \ac{CCMF} shows marginal evidence of a peak, the location of a peak is only a factor of $\sim 2$ higher than $M_\mathrm{min}/M_{J,0}=2.42\times 10^{-2}(N_\mathrm{core,res}/8)$ if we adopt $N_\mathrm{core,res} = 8$ in \cref{eq:Mmin_given_dx}, requiring some caution in interpretation.
If we relax the criterion for cores being resolved from $N_\mathrm{core,res} = 8$ to $4$,
the nominal resolved mass drops to $M_\mathrm{min}/M_{J,0}=1.21\times 10^{-2}$.
While the \ac{CCMF} peak position for model \texttt{M5} does not vary much after including these lower mass cores, the peak shifts toward lower mass for model \texttt{M10}.
We note, however, that at higher resolution, turbulent dissipation would have been lower in these less resolved cores and they would have been prevented from collapsing.

If cores were supported only by thermal pressure, one might expect a characteristic mass for collapse in a turbulent simulation to be given by the \ac{BE} mass in \cref{eq:mbe} when $\bar{\rho} $ is set equal to the characteristic post-shock value based on the large scale Mach number in the cloud,
\begin{equation}\label{eq:mchar}
    M_\mathrm{char,th} = 1.86 \frac{c_s^3}{G^{3/2}\rho_0^{1/2}\mathcal{M}_\mathrm{3D}} = 0.334\frac{M_{J,0}}{\mathcal{M}_\mathrm{3D}}
\end{equation}
\citep[e.g.][]{gong11,chen14,haugbolle18}.
Allowing for turbulent support of cores, this would be generalized to
\begin{equation}\label{eq:mcrit_ps}
  \begin{split}
    M_\mathrm{char,trb} &\equiv M_\mathrm{crit}(\overline{\rho} = \rho_0 \mathcal{M}_\mathrm{3D}^2)\\
    &\approx 0.334 \frac{M_{J,0}}{\mathcal{M}_\mathrm{3D}} \left(1 + \frac{1}{2}\frac{\sigma_\mathrm{1D}^2}{c_s^2}\right)
  \end{split}
\end{equation}
where $M_\mathrm{crit}$ is the critical \ac{TES} mass including turbulent support and we use \cref{eq:mcrit_fit} in the second line for $p=0.5$.
Taking $\sigma_\mathrm{1D} = 0.8c_s$ as a representative value in \cref{fig:tes_parameters}(d), \cref{eq:mcrit_ps} indicates that the characteristic mass including turbulent support is only slightly higher than the \ac{BE} mass at the post-shock density given above, $M_\mathrm{char,trb} = 1.32 M_\mathrm{char,th}$.
For $\mathcal{M}_\mathrm{3D} = 5$ and $10$, this corresponds to $M_\mathrm{char,trb} = 8.8\times 10^{-2}\,M_{J,0}$ and $4.4\times 10^{-2}\,M_{J,0}$, respectively, which are plotted in vertical shaded bands in \cref{fig:mass_function}(a).
For model \texttt{M5} where the peak of the \ac{CCMF} appears to be marginally resolved, the apparent peak occurs at $\sim 0.5M_\mathrm{char,trb}$.
Scaling from this result, the expected peak position for model \texttt{M10} would then be $0.5M_\mathrm{char,trb} \sim 2.2\times 10^{-2} M_{J,0}$.
This is close to \ac{CCMF} peak for the dashed pdf, which adopts $N_\mathrm{core,res}=4$.
However, we consider this too close to $M_\mathrm{min}=2.42\times 10^{-2} M_{J,0}(N_\mathrm{core,res}/8)$ to be properly resolved.
Both higher resolution simulations, possibly aided by adaptive mesh refinement techniques with careful refinement criteria, and extension of the parameter space to higher Mach numbers, would be required to convincingly address the questions of (1) whether or not there exists a numerically converged peak in the \ac{CCMF}, and (2) whether the peak is consistent with theoretical hypotheses (see \autoref{app:appendix} for related discussion).

\cref{fig:mass_function}(b) plots a similar distribution for core radius.
Again, the $\mathcal{M}_\mathrm{3D} = 5$ histogram suggests that the peak in the distribution occurs at half of the characteristic radius 
\begin{equation}
\label{eq:rcrit_ps}
  \begin{split}
    R_\mathrm{char,trb} &\equiv R_\mathrm{crit}(\overline{\rho} = \rho_0 \mathcal{M}_\mathrm{3D}^2)\\
    &\approx 0.430\frac{L_{J,0}}{\mathcal{M}_\mathrm{3D}} \left(1 + \frac{1}{2}\frac{\sigma_\mathrm{1D}^2}{c_s^2}\right)^{1/3}.
  \end{split}    
\end{equation}

\cref{fig:mass_function}(c) shows $M_\mathrm{core}$ is roughly a factor of $\sim 2$ higher than $M_\mathrm{BE}$ measured at the average density $\overline{\rho}$, although a few cores have even higher ratio.
While this enhancement can be attributed to turbulent support (i.e., $M_\mathrm{crit} \ge M_\mathrm{BE}$, as in \cref{eq:mcrit_fit}), some cores have additional mass excess due to rapid core building, as we will show later in \cref{fig:building_time}.

Finally, \cref{fig:mass_function}(d) plots the distribution of $R_\mathrm{core}/R_\mathrm{BE}$ which is equivalent to $r_\mathrm{crit}/R_\mathrm{BE}$ for critical cores.
It shows that the critical radius is larger than the \ac{BE} radius by no more than a factor of $\sim 2$, which is expected for transonic cores (\cref{eq:rcrit_fit}).

\begin{figure*}[htpb]
  \epsscale{1.18}
  \plotone{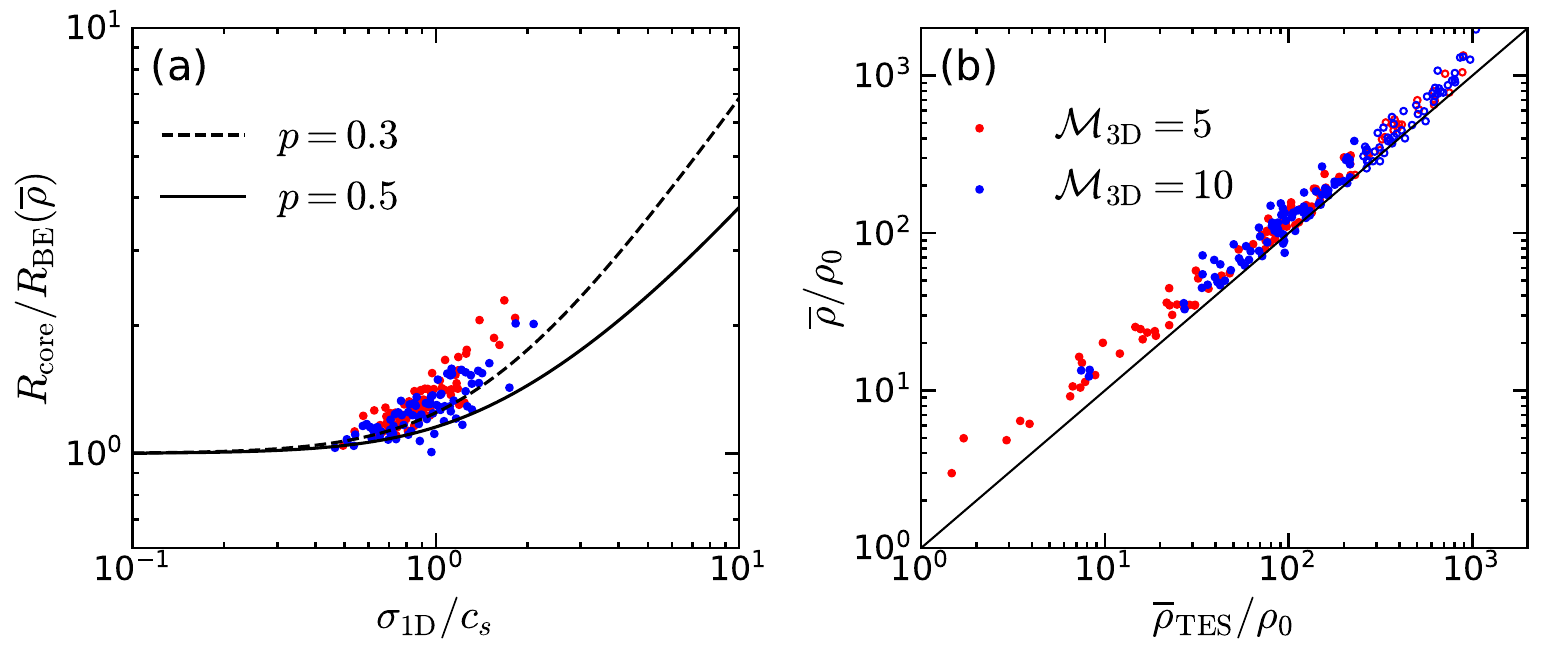}
  \caption{Comparison of measured core properties with \ac{TES} solutions, for the critical cores in models \texttt{M5} (red) and \texttt{M10} (blue). 
  (a) The ratio of the core radius to the \ac{BE} radius at the average density as a function of the turbulent Mach number.
  The solid and dashed lines correspond to the loci of \acp{TES} with $p=0.5$ and $0.3$, respectively.
  (b) The average density within $R_\mathrm{core}$ versus the predicted density of $p=0.5$ \acp{TES} (\cref{eq:rhotes}).
  Filled and open symbols denote critical cores satisfying the resolution criterion with $N_\mathrm{res,core}=8$ and $4$, respectively.}
  \label{fig:comparison_with_tes}
\end{figure*}

In \cref{fig:comparison_with_tes}(a), we plot the ratio $R_\mathrm{core}/R_\mathrm{BE}(\overline{\rho})$ as a function of the one-dimensional Mach number $\sigma_\mathrm{1D}/c_s$.
Overall, critical cores follow the curves defined by the \ac{TES} solutions, although the measured ratio is slightly ($\sim 10\%$) offset to higher values due to enhanced $\overline{\rho}$ compared to strict equilibrium (\cref{fig:comparison_with_tes}(b); we shall explain a potential reason for this in \cref{fig:building_time}).
It is worth noting that, even though the \ac{TES} model in principle extends to either a purely thermal regime with $\sigma_\mathrm{1D} \ll c_s$ or a highly turbulent regime with $\sigma_\mathrm{1D} \gg c_s$, all of the critical cores in our simulations are found in the regime $0.5 < \sigma_\mathrm{1D}/c_s < 2$.
We find no critical cores with either very low or very high levels of internal turbulence.
This is because 1) cores have a turbulent origin and are able to initiate collapse before $\sigma_\mathrm{1D}$ completely dissipates, so that there are no purely thermal cores; 2) highly turbulent cores would be self-destructive, in the sense of being torn apart before they can collapse.
In particular, Figure 8(d) of \citetalias{moon24} shows for \ac{TES} solutions that the flow crossing time is shorter than the average free-fall time for cores with $\sigma_\mathrm{1D}/c_s >2$.
As a result, the critical cores formed in our simulations are largely transonic, which is also likely the case for real cores.

\subsection{Core Shape Distributions}\label{sec:core_shapes}

\REV{
It is well known from observations \citep{myers91,ryden96,jones01,tassis07,lomax13} and from simulations both with \citep{gammie03,basu04,li04,nakamura08,chen18} and without \citep{klessenburkert00,offner09coreshape,gong11} magnetic fields that dense cores at various evolutionary stages are intrinsically triaxial in terms of their density distribution.
}

\begin{figure}[htpb]
  \epsscale{1.18}
  \plotone{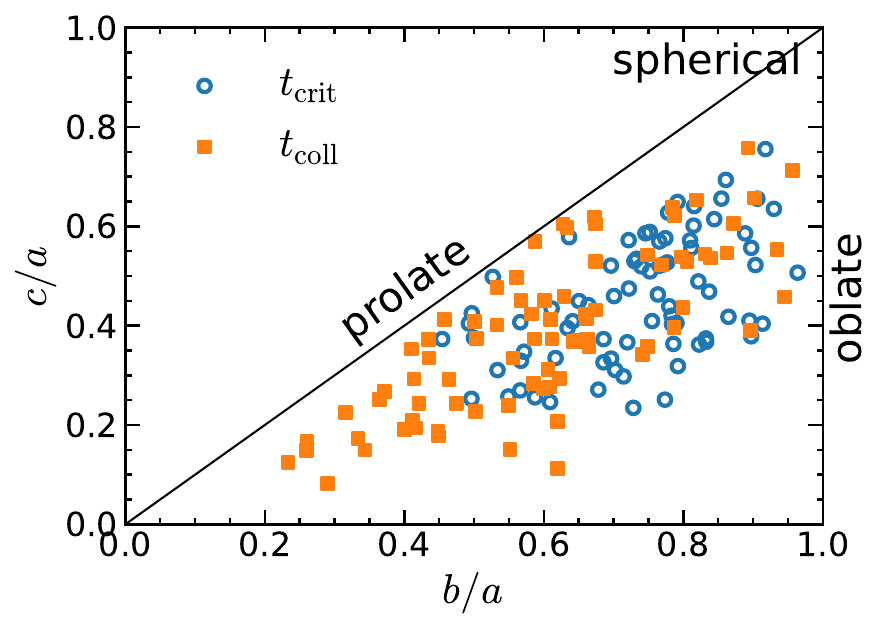}
  \caption{Axis ratio distribution for cores in model \texttt{M10} at $t_\mathrm{crit}$ (blue circles) and $t_\mathrm{coll}$ (orange squares). $a$, $b$, and $c$ are the three eigenvalues (largest to smallest) of the moment of inertia tensor calculated within $R_\mathrm{core}$.}
  \label{fig:shape_distribution}
\end{figure}

\REV{
In this section, we take advantage of our dynamical identification of $t_\mathrm{crit}$ (\cref{eq:tcritdef}) and $t_\mathrm{coll}$ (\cref{eq:tcolldef}) to characterize this shape distribution at the beginning and at the end of the collapse.
To do this, we calculate the moment of inertia tensor $I_{ij} \equiv \int \rho x_i x_j dV$ by integrating over a sphere of radius $R_\mathrm{core}$ centered at each core, and then find its three eigenvalues $a \ge b \ge c$.
\cref{fig:shape_distribution} plots the axis ratios $b/a$ and $c/a$ for cores in model \texttt{M10} satisfying $N_\mathrm{res,core} = 8$, taken at $t_\mathrm{crit}$ and $t_\mathrm{coll}$.
The mean and standard deviation for the axis ratios are $b/a = 0.73\pm 0.12$ and $c/a = 0.45\pm 0.12$ at $t_\mathrm{crit}$, and $b/a = 0.60\pm 0.18$ and $c/a = 0.39 \pm 0.16$ at $t_\mathrm{coll}$, indicating that the cores are generally triaxial at all times with a broad axis ratio distribution.
The overall smaller axis ratios at $t_\mathrm{coll}$ compared to at $t_\mathrm{crit}$ suggest that the cores become more elongated as collapse progresses, as expected for gravitational collapse \citep{lin65}.
}

\REV{
We note that the main body of the analysis presented in this work and in \citetalias{paperII} is based on angle-averaged radial profiles rather than density or gravitational potential contours.
The equation of motion for the angle-averaged radial velocity (Equation \eqlageom of \citetalias{paperII}) does not itself require the density distribution to be spherically symmetric.
The \ac{TES} model, does, however, involve additional assumptions that the mass-weighted mean rotational velocity is zero and that the mass-weighted turbulent velocities are statistically isotropic \citepalias{moon24}, which may be violated for triaxial cores due to anisotropic density weighting.
Nonetheless, the overall agreement between the \ac{TES} model and our simulated cores (e.g., \cref{fig:radial_profiles_tcrit,fig:comparison_with_tes}) suggests that the \ac{TES} model provides a reasonable approximation.
}

\section{Accretion Rates and Evolutionary Stages}\label{sec:accretion}

\begin{figure*}[htpb]
  \epsscale{1.18}
  \plotone{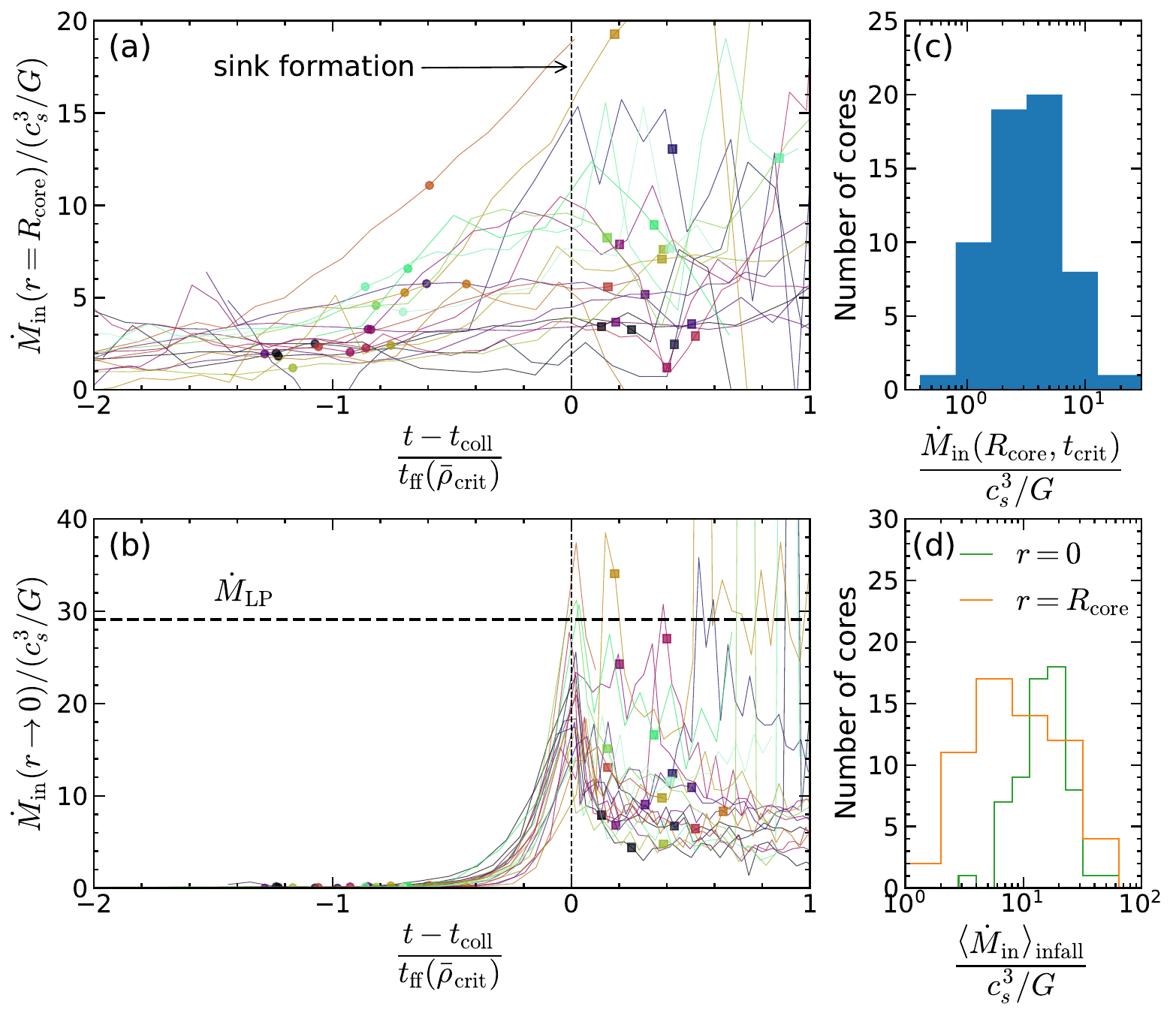}
  \caption{Accretion history and statistics.
  Left panels show temporal histories of the accretion rate for 20 randomly selected cores in model \texttt{M10}, measured at (a) $r = R_\mathrm{core}$ and (b) $r = 0$ (see the main text for definitions).
  For each core, time is measured from $t_\mathrm{coll}$ (when a sink particle forms), such that negative (positive) time corresponds to the prestellar (protostellar) stage, and is normalized by the free-fall time at the mean core density $\overline{\rho}_\mathrm{crit} \equiv 3M_\mathrm{core} / (4\pi R_\mathrm{core}^3)$.
  Circles and squares mark $t_\mathrm{crit}$ and $t_\mathrm{*,100}$ for each individual core, the latter of which is undefined for some sink particles that merge with another particle.
  The horizontal dashed line in panel (b) marks the asymptotic accretion rate of the \ac{LP} flow defined in \cref{eq:mdot_LP}.
  Right panels show accretion statistics for all cores in model \texttt{M10}.
  (c) The distribution of the inflow rate measured at $r=R_\mathrm{core}$ and $t=t_\mathrm{crit}$.
  (d) The distribution of the time-averaged accretion rates at $r=R_\mathrm{core}$ (orange histogram) and $r=0$ (green histogram).
  The average is taken during the interval $[t_\mathrm{coll}, t_{*,100}]$, except some cases where core tracking stopped before $t_{*,100}$.
  }
  \label{fig:accretion_rate}
\end{figure*}

\begin{figure*}[htpb]
  \epsscale{1.18}
  \plotone{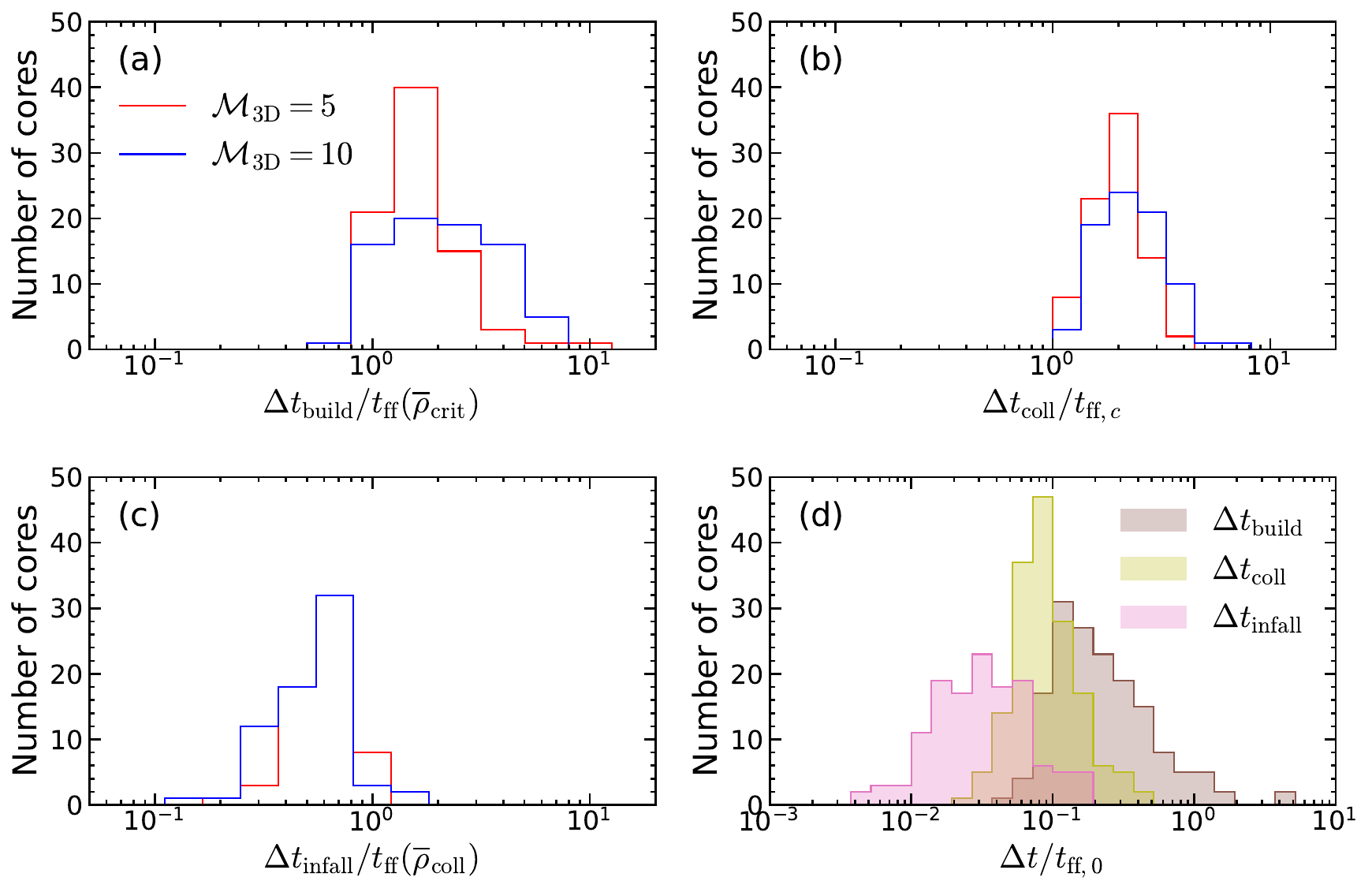}
  \caption{The distributions of the measured durations that a core spends in (a) the core building stage ($\Delta t_\mathrm{build}$), (b) the core collapse stage ($\Delta t_\mathrm{coll}$), and (c) the envelope infall stage ($\Delta t_\mathrm{infall}$), for cores in model \texttt{M5} (red) and \texttt{M10} (blue).
  The timescales are normalized by gravitational free-fall times at a density appropriate for each stage.
(d) The comparison of the combined distributions of $\Delta t_\mathrm{build}$ (brown), $\Delta t_\mathrm{coll}$ (yellow), and $\Delta t_\mathrm{infall}$ (pink) for all cores in models \texttt{M5} and \texttt{M10}; here the timescales are given in units of $t_\mathrm{ff,0}$, which is $\sim 3\text{--}6\,\mathrm{Myr}$ under typical \ac{GMC} conditions (\cref{eq:tff0}).}
  \label{fig:timescales}
\end{figure*}

In our simulations, cores form in regions where velocity fields are locally converging.
From a Eulerian perspective, the mass contained within a fixed radius $R_\mathrm{core}$ grows in time, reaching $M_\mathrm{core}$ at $t = t_\mathrm{crit}$ (see \cref{eq:Rcoredef,eq:Mcoredef} for definitions of $R_\mathrm{core}$ and $M_\mathrm{core}$).
To estimate the time taken to build up $M_\mathrm{core}$ within $R_\mathrm{core}$, we monitor the mass inflow rate
\begin{equation}\label{eq:mdot}
  \dot{M}_\mathrm{in} \equiv -\left<4\pi r^2\rho v_r\right> = -4\pi r^2 \left<\rho \right> \left<v_r \right>_\rho
\end{equation}
for each core.
\cref{fig:accretion_rate}(a) plots the time evolution of $\dot{M}_\mathrm{in}(r = R_\mathrm{core})$ for 20 randomly selected cores, showing that the mass accretion rate is roughly constant in time, although some cores show a slightly increasing trend.

Motivated by the limited variations in $\dot{M}_\mathrm{in}$ over time, we define the duration of the core building stage as
\begin{equation}\label{eq:building_time}
  \Delta t_\mathrm{build} \equiv \frac{M_\mathrm{core}}{\dot{M}_\mathrm{in}(r=R_\mathrm{core}, t=t_\mathrm{crit})}.
\end{equation}
\cref{fig:accretion_rate}(c) shows that the inflow rate responsible for the core building, $\dot{M}_\mathrm{in}(r=R_\mathrm{core}, t=t_\mathrm{crit})$, peaks at around $\sim 3 c_s^3/G$.
The time required to build a single \ac{BE} mass core is therefore $M_\mathrm{BE} / [3 c_s^3/G] \sim t_\mathrm{ff}$.
Because actual core mass at $t_\mathrm{crit}$ is generally larger than $M_\mathrm{BE}$ (\cref{fig:mass_function}(c)), the core building takes $\Delta t_\mathrm{build} \gtrsim t_\mathrm{ff}$ as shown in \cref{fig:timescales}(a).

\begin{figure}[htpb]
  \epsscale{1.18}
  \plotone{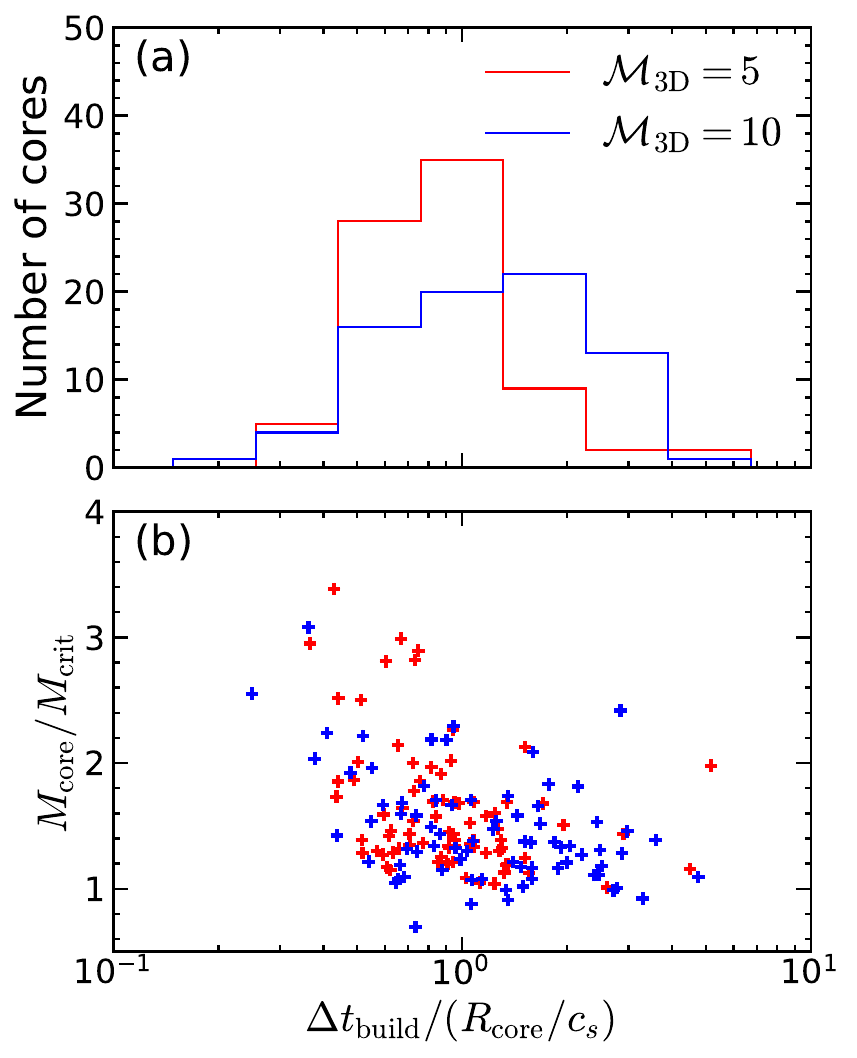}
  \caption{(a) The distribution of the ratio between the core building time (\cref{eq:building_time}) and the sound crossing time for model \texttt{M5} (red) and \texttt{M10} (blue). (b) The ratio between the core mass to the critical mass versus the normalized core building time.
  The critical mass is calculated using \cref{eq:mcrit_fit} with the measured average density $\overline{\rho}$ and velocity dispersion $\sigma_\mathrm{1D}$ of each critical core.
  The median ratio is $M_\mathrm{core}/M_\mathrm{crit} = 1.6$ and $1.3$ for the range $\Delta t_\mathrm{build}/(R_\mathrm{core}/c_s) < 1$ and $>1$, respectively.}
  \label{fig:building_time}
\end{figure}

\begin{figure}[htpb]
  \epsscale{1.18}
  \plotone{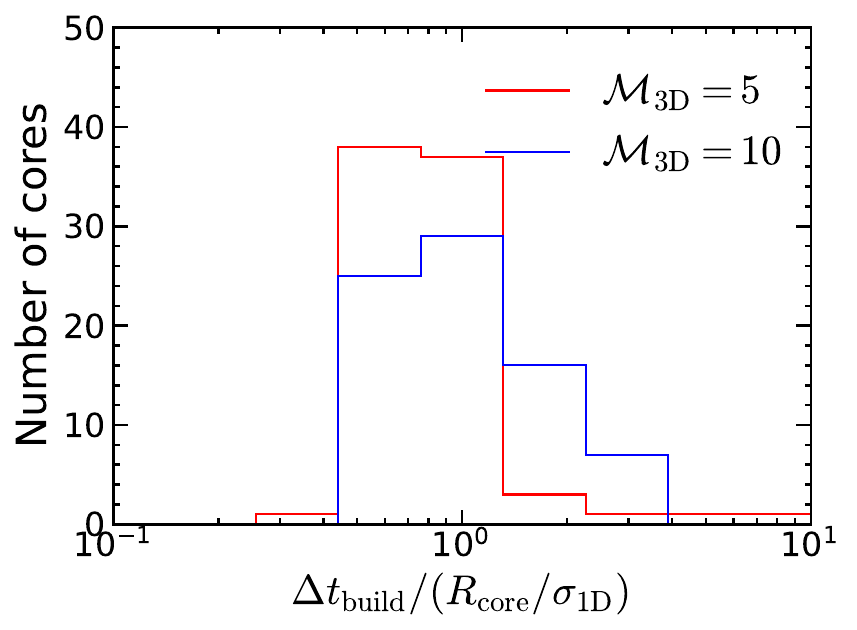}
  \caption{Similar to \cref{fig:building_time}(a), but for the building time normalized to the turbulent crossing time $R_\mathrm{core}/\sigma_\mathrm{1D}$.}
  \label{fig:building_time_trb}
\end{figure}

Also interesting is to compare the building time to the sound crossing time $R_\mathrm{core}/c_s$.
\cref{fig:building_time}(a) shows that most cores are built within a single sound crossing time, meaning that core building occurs dynamically rather than quasi-statically, consistent with a large scatter in the measured net force before the critical time (see Figure \figcoreevolution and \figpredictedforceevolution of \citetalias{paperII}).
Nonetheless, the core mass $M_\mathrm{core}$ measured at $t_\mathrm{crit}$ is comparable to the critical \ac{TES} mass $M_\mathrm{crit}$ within a factor of $\sim 2$ (see \cref{fig:building_time}(b)), indicating that most cores manage to approach a quasi-equilibrium at the end of the building stage (see also \cref{fig:radial_profiles_tcrit} for radial density profiles).
Interestingly, \cref{fig:building_time}(b) shows that the ratio $M_\mathrm{core}/M_\mathrm{crit}$ increases with decreasing building time for $\Delta t_\mathrm{build} < R_\mathrm{core}/c_s$, whereas for $\Delta t_\mathrm{build} > R_\mathrm{core}/c_s$ the ratio is relatively constant at $M_\mathrm{core} = 1.3 M_\mathrm{crit}$ (for comparison, the median ratio is $M_\mathrm{core}/M_\mathrm{crit} = 1.6$ for cores with $t_\mathrm{build} < R_\mathrm{core}/c_s$).
This suggests that the cores built rapidly by strong converging flows may never go through a quasi-equilibrium configuration before collapsing; however, such cases are rare in our simulations.
The dynamic nature of core building may partially explain why some cores show radial profiles deviating from the equilibrium solution (\cref{fig:radial_profiles_tcrit}).

To monitor the collapsing flow near the center and subsequent accretion onto the sink particle, we also measure the mass accretion rate at the center, $\dot{M}_\mathrm{in}(r \to 0)$.
We have used the notation $\dot{M}_\mathrm{in}(r \to 0)$ because, strictly speaking, the accretion rate must be formally zero at $r=0$.
In practice, we take $\dot{M}_\mathrm{in}(r\to 0)$ to be \cref{eq:mdot} evaluated at $r = \Delta x$ before a sink particle forms (i.e., $t < t_\mathrm{coll}$), whereas we use the sink particle accretion rate defined by
\begin{equation}
  \dot{M}_\mathrm{sink}(t) \equiv \frac{M_\mathrm{sink}(t) - M_\mathrm{sink}(t - 10^{-3}t_{J,0})}{10^{-3}t_{J,0}}
\end{equation}
to represent $\dot{M}_\mathrm{in}(r\to 0)$ after $t_\mathrm{coll}$.
\cref{fig:accretion_rate}(b) shows that $\dot{M}_\mathrm{in}(r\to 0)$ is essentially zero before $t_\mathrm{crit}$, while it rapidly increases as $t\to t_\mathrm{coll}$, approaching (but not reaching) the corresponding value of the \ac{LP} similarity solution,
\begin{equation}\label{eq:mdot_LP}
  \dot{M}_\mathrm{LP} = 29.1 \frac{c_s^3}{G},
\end{equation}
although our limited output time resolution prevents us from measuring $\dot{M}_\mathrm{in}(r\to 0)$ very close to $t_\mathrm{coll}$.
We note that the rapid increase of $\dot{M}_\mathrm{in}(r\to 0)$ does not occur immediately after $t_\mathrm{crit}$, indicating that it takes time for the wave of outside-in collapse to reach the center.

For a centrally concentrated sphere initially at rest, it takes a central free-fall time $t_{\mathrm{ff},c} \equiv [3\pi/(32G\rho_c)]^{1/2}$ to form a singularity at the origin when the pressure is negligible.
\cref{fig:timescales}(b) shows that the duration of the core collapse stage $\Delta t_\mathrm{coll}$ defined in \cref{eq:dtcoll} is narrowly distributed around the median $\Delta t_\mathrm{coll} = 2t_{\mathrm{ff},c}$, indicating that the collapse monitored in our simulations is two times slower than a pressureless free-fall.
We note that this is quantitatively consistent with the predicted collapse duration from the measured net force within cores (see Section \seccontrolledcollapse of \citetalias{paperII}).

After a sink particle forms at $t_\mathrm{coll}$, the accretion rate at the center quickly declines and remains roughly constant at a level a few times smaller than $\dot{M}_\mathrm{LP}$.
This is because only the very central part of the flow manages to reach the \ac{LP} asymptotic solution \citep{hunter77,foster93,gong09}.
Due to the continued accretion, the mass of a sink particle grows roughly linearly with time.

We define the time $t_\mathrm{*,100}$ as the instant at which $M_\mathrm{sink} = M_\mathrm{core}$.
This time is when all the progenitor core mass has fallen to the sink (i.e., core-to-star formation efficiency of $100\%$)\footnote{In reality, the accretion after $t_\mathrm{coll}$ would be affected by winds and radiation from a protostellar system, which are not included in our simulations.}.
The period of time between $t_\mathrm{coll}$ and $t_\mathrm{*,100}$ corresponds to the ``envelope infall'' stage of \citet{gong09}, for which we define
\begin{equation}
  \Delta t_\mathrm{infall} \equiv t_\mathrm{*,100} - t_\mathrm{coll}.
\end{equation}
\cref{fig:timescales}(c) shows that $\Delta t_\mathrm{infall}$ is typically shorter than the free-fall time associated with the mean core density $\overline{\rho}_\mathrm{coll}$ at the end of the collapse\footnote{As the collapse proceeds the mean density within the ``Lagrangian radius'' $r_M$ enclosing $M_\mathrm{core}$ increases such that $\overline{\rho}_\mathrm{coll} \equiv M_\mathrm{core}/(4\pi r_M^3/3) \sim 3 \overline{\rho}_\mathrm{crit}$, where $\overline{\rho}_\mathrm{crit} \equiv M_\mathrm{core}/(4\pi R_\mathrm{core}^3/3)$.
The free-fall time correspondingly decreases to $t_\mathrm{ff}(\overline{\rho}_\mathrm{coll}) \sim 0.6 t_\mathrm{ff}(\overline{\rho}_\mathrm{crit}).$}, with the median at $\Delta t_\mathrm{infall} = 0.6 t_\mathrm{ff}(\overline{\rho}_\mathrm{coll})$.
This is not surprising considering that by the time $t_\mathrm{coll}$, the core inflow has already achieved supersonic velocities (see Figure \figcoreevolution in \citetalias{paperII}) such that it takes less time to reach the center compared to the free-fall from initially at rest.
We note that \citet[Figure 12]{gong09} found $\Delta t_\mathrm{infall}/t_\mathrm{ff}(\overline{\rho}_\mathrm{coll}) \sim 0.8\text{--}1$ in their one-dimensional converging flow simulations.

\cref{fig:accretion_rate}(b) shows that the sink mass continues to grow even after $t_\mathrm{*,100}$ without noticeable decline in the accretion rate.
As previously pointed out by \citet[see their Figure 20]{gong15}, this late accretion \citep[in the terminology of][]{gong09} is expected since cores are not isolated objects truncated at a finite radius, but instead are the inner parts of converging flows that continue to be accreted onto the sink particle.
For example, \cref{fig:accretion_rate}(d) indicates that the time-averaged inflow rate at $r = R_\mathrm{core}$ during the envelope infall stage is about a half as large as the sink particle accretion rate during the same time period.

In reality, however, the accreting flows likely arrive at a protostellar disk rather than directly onto the forming protostar, such that only a fraction of $\dot{M}_\mathrm{in}(r\to 0)$ would ultimately reach the protostellar surface.
In addition, radiation and/or bipolar outflows from the forming protostar would limit the accretion rate during the envelope infall and late accretion stages for some systems \citep[e.g.,][]{hansen12,kuiper16}.
However, the persistence of accretion after the initial core is consumed may be essential for the formation of massive stars.
Based on our simulations, it does not appear possible for massive stars to originate as massive, highly turbulent, quasi-equilibrium cores that become supercritical and collapse.
An alternative would be sustained accretion from a larger-scale converging flow.

Finally, \cref{fig:timescales}(d) compares $\Delta t_\mathrm{build}$, $\Delta t_\mathrm{coll}$, and $\Delta t_\mathrm{infall}$ on a common scale, showing that the characteristic timescales becoming shorter and shorter as a core evolves to later stages.
The median value for each stages are $\Delta t_\mathrm{build} \sim 0.19 t_\mathrm{ff,0}$, $\Delta t_\mathrm{coll} \sim 0.086 t_\mathrm{ff,0}$, and $\Delta t_\mathrm{infall} \sim 0.031 t_\mathrm{ff,0}$, where $t_\mathrm{ff,0} \sim 3\text{--}6\,\mathrm{Myr}$ under typical \ac{GMC} conditions (\cref{eq:tff0}).

\section{Discussion}\label{sec:discussion}

\subsection{Is There a Threshold Density?}\label{sec:discussion_collapse_condition}

A recurring theme in star formation theory is that there exists a ``threshold density'' above which gas is gravitationally unstable.
For example, based on different physical arguments and assumptions, \citet{krumholz05} and \citet{padoan11} proposed that there exists a threshold density above which a core becomes unstable and collapses.
For a \REV{spherical} cloud with the virial parameter $\alpha_\mathrm{vir} = 5\sigma_\mathrm{1D}^2 R_\mathrm{cloud}/(GM_\mathrm{cloud})$ and Mach number $\mathcal{M}_\mathrm{3D} = \sqrt{3}\sigma_\mathrm{1D}/c_s$, the threshold density (relative to the mean cloud value $\rho_0$) predicted from these theories is $\rho_\mathrm{thr}/\rho_0 = (0.28\text{--}0.55)\times \alpha_\mathrm{vir}\mathcal{M}_\mathrm{3D}^2$ (the smaller and larger coefficients correspond to the theory of \citealt{krumholz05} and \citealt{padoan11}, respectively, with their fiducial choice of order-unity parameters; we assume $p=0.5$ for the former theory).

When applied to our models \REV{by taking $M_\mathrm{cloud} = M_\mathrm{box}$ and $4\pi R_\mathrm{cloud}^3/3 = L_\mathrm{box}^3$ leading to $\alpha_\mathrm{vir} \approx 2$}, the theories of \citet{krumholz05} and \citet{padoan11} predicts that the threshold density lies somewhere in between $\rho_\mathrm{thr}/\rho_0 = 14\text{--}28$ and $56\text{--}110$ for models \texttt{M5} and \texttt{M10}, respectively.
Although the measured distributions of critical core \REV{mean} density (\cref{fig:core_densities}(c)) start to rise roughly around the predicted range of $\rho_\mathrm{thr}$, the distribution extends widely above and below $\rho_\mathrm{thr}$.
This indicates that the threshold density of \citet{krumholz05} or \citet{padoan11} is neither a necessary nor sufficient condition for collapse of individual cores. \REV{\cref{fig:core_densities}(a) also shows a broad range of central density at the critical time.}
Indeed, \cref{fig:rs_rhoc_correlation} indicates that individual cores start to collapse at a wide range of densities, due both to the spatially varying strength of turbulence (\cref{fig:density_velocity_profiles}(a)), and to the varying tidal environment in which they are born.

Rather than a single threshold density, there may be a characteristic density set by the global parameters of a cloud, with more specific critical criteria responsive to local variations in conditions.  
We note that very different physical arguments lead to similar characteristic densities of gravitationally unstable cores.
For example, considerations of anisotropic core formation mediated by magnetic fields in local converging flows lead to $\rho_\mathrm{char}/\rho_0 = 0.56\mathcal{M}_\mathrm{3D}^2$ for ${\cal M}_\mathrm{3D}$ the flow Mach number \citep[from Eqn. (6) and (7); see also \citealt{chen14}]{chen15}.
For $\alpha_\mathrm{vir} \sim 2$, all these characteristic densities are comparable to a simple ``post-shock'' density $\mathcal{M}_\mathrm{3D}^2\rho_0$ multiplied by some order unity coefficient.

The \ac{TES} model provides a more mathematical route to a characteristic unstable-core density based on cloud-scale parameters, yielding a similar value to characteristic densities obtained from various physical arguments described above.
Combining \cref{eq:rhotes} with \cref{eq:sigrcrs} (assuming $p=0.5$ and $\eta_d=0.9$) and \cref{eq:rs_cloud}, the average density of critical cores forming \REV{in a spherical cloud} under ``typical'' turbulence conditions (i.e., $r_s \sim r_{s,\mathrm{cloud}} = (9/4)(R_\mathrm{cloud}/\mathcal{M}_\mathrm{3D}^2)$) is
\begin{equation}\label{eq:rhochar_from_tes}
\begin{split}
   \frac{\rho_\mathrm{char}}{\rho_0} &= 0.11 
   \frac{\left[1 + \frac{1}{2}\left(\frac{\sigma_\mathrm{1D}}{c_s}\right)^{2}\right]^{2/3}}{\left(\frac{\sigma_\mathrm{1D}}{c_s}\right)^{4}}\alpha_\mathrm{vir}\mathcal{M}_\mathrm{3D}^2
   \\
    &\sim 0.3\alpha_\mathrm{vir}\mathcal{M}_\mathrm{3D}^2,
\end{split}
\end{equation}
where we take $\sigma_\mathrm{1D} = 0.8c_s$ as a representative value (e.g., \cref{fig:tes_parameters}(d)) in the second equality.
For a cloud with $\alpha_\mathrm{vir} \sim 2\text{--}4$, \cref{eq:rhochar_from_tes} says that $\rho_\mathrm{char}$ is essentially the post-shock density, $\mathcal{M}_\mathrm{3D}^2\rho_0$, for an isothermal shock.

Of course, in order for collapse to occur, the core radius -- which depends on the local tidal gravitational field -- must exceed the critical radius.  If we use  the above $\rho_\mathrm{char}$ in \cref{eq:rcrit_fit} and \cref{eq:mcrit_fit}, adopting $\sigma_\mathrm{1D}/c_s =0.8$ as before, we find a characteristic radius and mass for critical cores of $R_\mathrm{char}=0.83 L_{\mathrm{J},0}/(\alpha_\mathrm{vir}^{1/2}\mathcal{M}_\mathrm{3D})$ \REV{(comparable to the mean sonic radius in the cloud, from \cref{eq:rs_cloud})} and $M_\mathrm{char}=0.78 M_{\mathrm{J},0}/(\alpha_\mathrm{vir}^{1/2}\mathcal{M}_\mathrm{3D})$, where \cref{eq:ljeans} and \cref{eq:mjeans} can be used to convert to physical units.
For typical \ac{GMC} parameters, $R_\mathrm{char} \sim 0.08\  {\mathrm pc}$ and $M_\mathrm{char} \sim 1.8 M_\odot$.

When considering the effects of turbulence, almost all theories assume that a single linewidth--size relation applies to all cores and that the turbulent velocities are completely independent of the density.
However, \cref{fig:density_velocity_profiles}(a) (see also Figure \figlinewidthsize of \citetalias{paperII}) shows that both the slope and normalization of the \emph{local} linewidth--size relations significantly vary from region to region.
While we took $r_s = r_{s,\mathrm{cloud}}$ and $p=0.5$ in deriving \cref{eq:rhochar_from_tes}, in reality, these local variations in the turbulent scaling relations would modify the leading numerical coefficient in \cref{eq:rhochar_from_tes} proportional to 
$(r_s/r_{s,\mathrm{cloud}})^{-2}\propto (\sigma_\mathrm{1D}/c_s)^4$, in addition to the variation in  $\sigma_\mathrm{1D}/c_s$ of the square bracketed factors.
This would lead to a range of critical densities at which collapse is triggered, as is seen in our distributions in \cref{fig:core_densities} (see also \cref{fig:rs_rhoc_correlation}).
A more refined theory should therefore draw the turbulent velocity dispersion (or the local sonic scale) from a joint distribution of density and velocity dispersion rather than assuming a single linewidth--size relation independent of density.

\REV{To summarize, \cref{eq:rhochar_from_tes} implies that the variation in the characteristic density $\rho_\mathrm{char}$ reflects differences in global physical conditions --- $\rho_0$, $\alpha_\mathrm{vir}$, and $\mathcal{M}_\mathrm{3D}$ --- across star forming regions.
In addition to this, within a single cloud, correlated variations between local density and turbulence result in a distribution of critical densities around $\rho_\mathrm{char}$ (see \cref{fig:density_velocity_profiles}).
}

\subsection{Absence of highly-turbulent cores}\label{sec:massive_cores}

\begin{figure}[htpb]
  \epsscale{1.18}
  \plotone{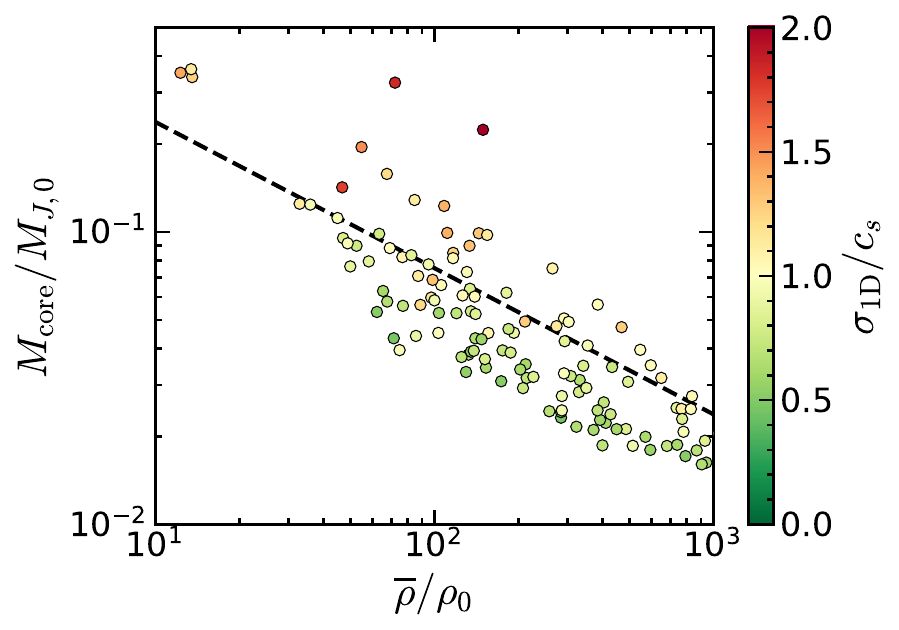}
  \caption{\REV{Core mass versus mean core density for the critical cores in model \texttt{M10} satisfying $r_\mathrm{crit} \ge 4\Delta x$ (i.e., $N_\mathrm{core,res} = 4$), color coded by their internal velocity dispersion.
  Dashed line plots
  $M_\mathrm{core} = 1.5M_\mathrm{crit}$ for $M_\mathrm{crit}$ as given in \cref{eq:mcrit_fit} adopting $\sigma_\mathrm{1D} = c_s$, 
  with the factor $1.5$ the typical mass excess found in our simulations (\cref{fig:building_time}(b)). As predicted in \cref{eq:mcrit_fit}, core mass also increases with core velocity dispersion (green to red).  
 }
  }
  \label{fig:mcore_vs_rho}
\end{figure}

Our simulations show that the vast majority (96\%) of the critical cores that are the immediate precursors of star formation have radial turbulent velocity dispersion $\sigma_\mathrm{1D}<1.5c_s$, and even the most turbulent critical core has $\sigma_\mathrm{1D} = 2.1 c_s$.
That is, the turbulence is at most trans-sonic.
\REV{It is worth emphasizing that this lack of highly turbulent critical cores was not a foregone conclusion.
Indeed, the total mass $M_\mathrm{box} = 64M_{J,0}$ and turbulent velocity dispersion $\mathcal{M}_\mathrm{3D} = 10$ in model \texttt{M10} are large enough such that supersonically turbulent cores with $\sigma_\mathrm{1D}/c_s \gtrsim 3$ could have formed.
For example, setting $\sigma_\mathrm{1D}/c_s = 5$ in \cref{eq:mcrit_fit} yields $M_\mathrm{crit} = 4.5 (\bar\rho/\rho_0)^{-1/2} M_{J,0} \ll M_\mathrm{box}$ for mean core density $\bar\rho$.
Our results therefore suggest that, although the \ac{TES} theory and the initial condition of our numerical models permit the existence of highly turbulent cores, they are not easily realized in the actual outcome.
\cref{eq:mcrit_fit} suggests that core mass is inversely correlated with local density and positively correlated with local velocity dispersion, and our numerical results are consistent with this (see   \cref{fig:mcore_vs_rho}).  However, we find only a limited range of core velocity dispersion.
}

\RREV{In terms of mass, the \ac{CCMF} above the peak in our simulations is well fit by a lognormal function with the standard deviation $0.55$, consistent with the Chabrier IMF.
It is notable also that (1) from \cref{fig:tes_parameters}(d), all of
the cores have relatively low internal turbulence levels (the distribution of $\sigma_\mathrm{1D}/c_s$ peaks at a subsonic level), and (2) the most massive cores in our simulations are not highly-turbulent cores, but rather those forming in lower density environments.
Purely hypothetically, our simulations could have produced quite different outcomes.
For example, only a few highly turbulent, very massive cores could have been produced.
The \ac{CCMF} in such a
hypothetical scenario would be very different from what we actually found and present in this paper, in terms of the peak position and the width of the distribution.}
\RREV{Taken together}, we do not find evidence supporting the \citet{mckee03} theory (see also \citealt{myers92,mclaughlin97}), in which the precursors of massive stars are massive cores supported by \REV{high-amplitude} turbulence\RREV{, although we cannot completely rule out the possible existence of such massive turbulent cores given our limited statistics and box-scale Mach number.}

From a physical point of view, the idea of massive cores supported primarily by turbulence suffers from two main difficulties.
One is that highly supersonic turbulence would produce a high degree of substructure, so that even if a core has strong enough gravity to collapse, it would rapidly fragment rather than producing a single massive star.
The second is that strong turbulence implies a flow crossing time comparable to the gravitational timescale, so that highly turbulent structures are also highly transient.
For critical \acp{TES}, the ratio of the turbulent crossing time $\Delta t_\mathrm{cross} \equiv R_\mathrm{core}/\sigma_\mathrm{1D}$ to the free-fall time at average density decreases from $\Delta t_\mathrm{cross}/t_\mathrm{ff} = 1.6$ at $\sigma_\mathrm{1D}/c_s = 1$ to $\Delta t_\mathrm{cross}/t_\mathrm{ff} = 0.82$ at $\sigma_\mathrm{1D}/c_s = 3$ (see also Figure 8(d) of \citetalias{moon24}).
\cref{fig:building_time_trb} indicates that \REV{on average the median core building time is shorter than the turbulent crossing time}, and that less than $10\%$ of cores have $\Delta t_\mathrm{build}/\Delta t_\mathrm{cross} > 2$.
This suggests that massive, turbulently supported cores would self-destruct before they succeed in assembling enough mass to become unstable.

Recent high-resolution observations also do not find strong evidence of high-mass, highly turbulent prestellar cores \citep[e.g.,][]{sanhueza19,redaelli22,li23,morii23,nony23}; often, previous high-mass prestellar core candidates turn out to be protostellar in nature or resolved into a collection of lower-mass substructures when resolution is increased.
Observed \emph{protostellar} cores in high-mass star-forming regions are on average more massive than prestellar cores and have shallower \ac{CMF} slope \citep{li23,nony23}, suggesting cores continue to grow in mass after collapse.
Taken together, the lack of \REV{highly turbulent,} massive prestellar cores and the presence of continued accretion well beyond the time of collapse (e.g., \cref{fig:accretion_rate}) support the idea that massive stars form by accretion flows from scales beyond the critical core (see Section 2.6.2 of \citealt{motte18} for a review; see also \citealt{padoan20}), rather than by collapse of highly turbulent, massive prestellar cores.

\subsection{The CMF in Numerical Simulations}\label{sec:cmf}

Various theoretical arguments suggest that as a result of supersonic turbulence, the \ac{CMF} defined by the number of cores per logarithmic mass bin has peak at a  characteristic mass scale set by the large-scale physical conditions in the cloud (e.g., \cref{eq:mchar,eq:mcrit_ps}, \citealt{gong11,chen14,chen15,haugbolle18}; see also \citealt{padoan02,hennebelle08,hopkins12} for 
related theories that make predictions for the whole CMF).
Even if the full \ac{CMF} does not directly map to the stellar \ac{IMF} due to late-stage accretion and disk fragmentation, a peak in the \ac{CMF} could potentially have implications for the turnover of the \ac{IMF} or the mass function of stellar systems, allowing for fragmentation into a binary or multiple at a later evolutionary stage \citep[see e.g. review by][]{2023ASPC..534..275O}.  

It has been questioned in the literature, however, whether cloud-scale conditions (including an effectively isothermal equation of state, given the balance of heating and cooling) and the physics of self-gravitating turbulence are ever able to imprint a characteristic mass scale, and if so, whether this is relevant to the \ac{IMF} \citep[e.g.,][see also Section 5 of \citealt{hennebelle24} for a recent review]{martel06,federrath17,guszejnov18,guszejnov20}. 
Regarding the first question, it has been suggested that non-convergence of mass functions is a general feature of self-gravitating, isothermal numerical simulations. For example, \citet{guszejnov18} found no convergence with increasing numerical resolution in the low-mass spectrum of the \ac{SMF} for their isothermal, unmagnetized self-gravitating turbulence simulations, 
and   
\citet[see their Figure 10]{guszejnov20} found lack of convergence in the low mass end of the \ac{SMF} in their isothermal \ac{MHD} simulations. 
Although their specific results regarding lack of low-mass convergence may have been compromised by an error in their sink particle implementation
\citep[see][]{guszejnov21}, a more general issue is that 
this approach does not distinguish between convergence in the \ac{CMF} and convergence in the \ac{SMF}.  
  
Indeed, when discussing numerical convergence of mass functions, most studies have presented results based on the \ac{SMF} rather than the \ac{CMF} (e.g., \citealt{martel06,federrath17,guszejnov18,lee18II,haugbolle18}; see, however, \citealt{chen14,chen15,gong15,padoan20,pelkonen21} for the \ac{CMF}).
While the \ac{CMF} in observations reflects cloud-scale dynamics and choices in segmentation algorithms, and the \ac{CCMF} studied in the present work focuses on the initial conditions for formation of star-disk systems, additional effects (both physical and numerical) enter in determining the \ac{SMF} in simulations, so that one does not necessarily map to the other. Collapsing cores or disks may fragment (although the fragmentation \emph{during} core collapse may be difficult; see \citealt{tohline80a,tohline80b,hanawa99,lai00,sugimura17}), and cores that collapse may continue to accrete, growing in mass (as shown by \cref{fig:accretion_rate}; see also \citealt{padoan20,pelkonen21}).
The \ac{SMF} measured at any given moment will include young sink particles that have accreted only a small fraction of the parental core mass, as well as old sink particles that become overmassive through sustained accretion if realistic feedback is not included.
Additionally, the \ac{SMF} may include spurious sink particles formed in unresolved cores that would otherwise have been torn apart by turbulence at higher resolution, and low-mass sinks forming within disks around sink particles at late stages of evolution.
Disk fragmentation is subject to tidal forces and heating once a central protostar has formed \citep[see][for a review]{kratter16}, and while these effects are important in setting the \ac{IMF} they are not relevant to the cloud-scale gravo-turbulent fragmentation process.
The \ac{SMF} can also be affected by the choice of subgrid model parameters, as shown by \citet[Appendix C]{haugbolle18}.
Thus, in order to properly distinguish the roles of gravo-turbulent fragmentation on cloud scales from disk-scale fragmentation, we strongly encourage measuring and studying \acp{CMF} in numerical simulations.
Although there is no single ``correct'' definition of a prestellar core, the criteria we have developed based on the theoretical critical radius compared to the tidal radius (and transition from positive to negative net force) provide physically motivated definitions, which leads to the \ac{CCMF} in this work.
Following the full history prior to, during, and subsequent to collapse would be necessary to test theories of the \ac{CMF} and to understand the relation between the \ac{CMF} and the \ac{IMF}.

To claim the existence or absence of a numerically converged mass function (whether it is the \ac{CMF} or the \ac{SMF}), it is necessary to achieve numerical resolution high enough to unambiguously resolve the anticipated characteristic mass.
In \autoref{app:appendix}, we intercompare various measures of numerical resolution in recent self-gravitating turbulence simulations, which is relevant to the question of numerical convergence in mass functions.

\subsection{Connection to Observations}

\REV{Because a primary goal in this work was to test the applicability of the semianalytic \ac{TES} model, the simulations presented in \citetalias{paperII} and in this work are highly idealized by construction.
Furthermore,  the physical properties of cores are measured at a very particular epoch, $t_\mathrm{crit}$, corresponding to the onset of runaway collapse for each core.
While it is therefore premature to attempt comprehensive comparisons to observed core properties, it is worthwhile noting a few key points.
Taking the reference values for the temperature and ambient density in \cref{eq:ljeans,eq:mjeans}, the typical mass and radius of the critical cores (\cref{fig:mass_function}) are, in physical units, $\sim 2\,M_\odot$ and $\sim 0.1\,\mathrm{pc}$, respectively, which are similar to the upper envelope of the observed starless/prestellar core distribution from Herschel/JCMT Gould Belt Surveys \citep{kirk13,konyves15,pattle15,salji15,kirk16,ward-thompson16,mowat17,pattle17,benedettini18,bresnahan18,difrancesco20,konyves20,fiorellino21,pezzuto21,kirk24,pattle25}.
We note that the critical core properties are taken at the onset of collapse; as the cores subsequently evolve to higher central column densities, their ``observational radius'' based on the column density FWHM would decrease over time.
The observed core radius also depends on adopted definitions and details of background subtraction procedure.
Hence, care must be taken to uniformly compare the observed and simulated core sample.}

\REV{
Observations generally report outer density profiles that are shallower than $r^{-2}$ but steeper than $r^{-1}$ \citep{bergin07,difrancesco07,chen19}.
This is similar to the theoretical prediction from the \ac{TES} model \citepalias[see Figures 6 and 7 there]{moon24} and to the density profiles of critical cores shown here (\cref{fig:radial_profiles_tcrit}).
Consistent with our findings for the critical cores in our simulations, spectral line surveys find that the average line widths of cores are largely transonic  \citep{cwlee99,cwlee01,foster09,friesen09,friesen10,lee14,storm16,keown17,kirk17,tang18,chen19,chung19,kerr19,choudhury21,chung21,li23,yoo23}.
}

\RREV{We also note that $M_\mathrm{core} \propto \overline{\rho}^{-1/2}$ relation evident in \cref{fig:mcore_vs_rho} suggests that the mass of critical cores is overall proportional to their size. In fact, combination of \cref{eq:rbe,eq:mbe,eq:rcrit_fit,eq:mcrit_fit} gives a mass-size relation for the critical \ac{TES}:
\begin{equation}\label{eq:mass_size_tes}
    M_\mathrm{crit} = 2.44 \frac{c_s^2}{G} r_\mathrm{crit}\left( 1 + \frac{1}{2}\frac{\sigma_\mathrm{1D}^2}{c_s^2}\right)^{2/3};
\end{equation}
for a limited range of $\sigma_\mathrm{1D}/c_s$ this would indeed represent a roughly linear relationship. 
In principle, \cref{eq:mass_size_tes} is of value when interpreting the observed mass--size relations of starless cores \citep[e.g.,][]{kirk13,konyves15,pattle15,salji15,kirk16,ward-thompson16,keown17,kirk17,mowat17,pattle17,benedettini18,bresnahan18,chen19,difrancesco20,konyves20,fiorellino21,pezzuto21,kirk24}. However, a direct comparison may be difficult, because observed cores are at various evolutionary stages (as opposed to at $t_\mathrm{crit}$; see however \citealt{simpson11} and \citealt{bresnahan18} for evolutionary modeling), and widely different methods are empirically adopted for defining core boundaries (and background subtracting), which affect both the mass and the size.
}

\REV{As already discussed in Section 6.2 of \citetalias{moon24}, it will be very interesting to examine how projection effects map the intrinsic power-law structure function found within the critical cores (\cref{fig:density_velocity_profiles}(a)) to the observed linewidth--size relation.
Our preliminary analysis (Moon \& Ostriker, \emph{in prep.}) indicates that, even though the intrinsic turbulent velocity dispersion grows in radius as a power law within cores, the line-of-sight velocity dispersion flattens toward the center in the sky plane due to density-weighted projection effects (see also Section 3.1.2 of \citealt{mckee07}).
This implies that observed ``coherent cores,'' which have roughly constant non-thermal velocity dispersion profiles and shallower-than-\ac{BE} density profiles \citep[e.g.,][]{goodman98,pineda10,chen19} might be partly supported by turbulence, and evolving toward the onset of collapse.}

\section{Summary and Conclusions}\label{sec:conclusion}

Prestellar cores form within turbulent \acp{GMC} and undergo gravitational runaway at some point in their life, leading to star formation.
To understand the physical processes responsible for triggering collapse, it is essential to investigate the detailed evolution of individual cores and their physical properties at the onset of collapse.
In the companion paper \citepalias{paperII}, we have conducted a comprehensive analysis of the evolution of individual cores, using a carefully constructed suite of numerical simulations of turbulent, self-gravitating clouds.
In this work, we investigate the structure and physical properties of these cores measured at the \emph{critical time} when they initiate collapse (\cref{sec:critical_core_properties}), and interpret these results using our new model for \acp{TES} developed in \citetalias{moon24}.
We also measure the inflow rates before and after the collapse, which determine the timescales the cores spend in each evolutionary stage (\cref{sec:accretion}).

Our main conclusions are as follows:
\begin{enumerate}

  \item
    Within cores, the turbulent velocity dispersion increases with distance $r$ approximately as a power-law (\cref{fig:density_velocity_profiles}(a)), validating the main assumption of the \ac{TES} model developed in \citetalias{moon24}.
    The measured \REV{angle-averaged} radial density profiles are \REV{close to} the theoretical prediction of the \ac{TES} model for both quiescent and highly turbulent cores (\cref{fig:radial_profiles_tcrit})\REV{, even though the majority of cores are triaxial with mean axis ratios of $0.73$ and $0.45$ (\cref{fig:shape_distribution}).}
    The \ac{BE} sphere profile is too steep to explain the density structure of turbulent cores.

  \item
    The cores in our simulations have subsonic or transonic internal velocity dispersions ($\sigma_\mathrm{1D} \sim 0.5\text{--}1.5 c_s$) at the onset of the collapse (\cref{fig:tes_parameters}(d)), consistent with observed superthermal linewidths \citep[e.g.,][]{cwlee99,cwlee01,foster09,friesen09,friesen10,lee14,storm16,keown17,kirk17,tang18,chen19,chung19,kerr19,choudhury21,chung21,li23,yoo23}.
    The slope of the internal linewidth--size relation of cores is $p \sim 0.5\pm 0.2$ (\cref{fig:tes_parameters}(a)), similar to the slope of the linewidth--size relation on large scales in our simulations (and observed \acp{GMC}).
    We do not find evidence for highly turbulent, massive prestellar cores (see \cref{sec:massive_cores} for related discussion).
    However, we do find evidence for continued accretion beyond the core collapse (\cref{fig:accretion_rate}(b)), which may be responsible for the formation of massive stars.

  \item
    The building, collapse, and infall stages of core evolution have successively shorter durations (\cref{fig:timescales}), decreasing from a median of $0.19$ to $0.086$ to $0.031$ times the free-fall time $t_\mathrm{ff,0}$ at the mean density of the surrounding cloud.
    This implies that observed timescales would become shorter as cores evolve from prestellar to protostellar phases.
    Since the mean core densities are $\sim 100$ times average cloud values, the duration of prestellar evolutionary stages is $\sim 3$ times the internal free-fall time within the critical cores.
    Core building takes only a single sound crossing time (\cref{fig:building_time}), indicating that cores form rapidly by dynamical compression rather than quasi-static processes.
    The prolonged collapse duration relative to the central free-fall time, $\Delta t_\mathrm{coll} \sim 2 t_{\mathrm{ff},c}$, is consistent with the small net force applied throughout the collapse \citepalias{paperII}.

  \item
    Cores form at a wide range of densities (\cref{fig:core_densities}), with most having a center-to-edge density contrast of $\sim 10\text{--}30$ (here, ``edge'' means $r=r_\mathrm{crit}$), consistent with theoretical expectations based on the \ac{TES} model in \citetalias{moon24} with at most transonic internal turbulence.
    The distribution of ratios of the core internal density to the mean cloud density on large scales peaks at $\bar{\rho}/\rho_0\sim 100$ for both our \texttt{M5} and \texttt{M10} models  (although the peak would likely shift to higher densities for model \texttt{M10} when resolution is increased),   
    extending down to $\sim 10$ for \texttt{M10}, and lower for \texttt{M5}.
    There is no sign of a single threshold density for collapse, as has been proposed in some theories for determining the star formation rate in molecular clouds.
    Local variations in the turbulence strength are significant even within a single cloud, affecting the density at which cores become unstable (\cref{fig:density_velocity_profiles,fig:rs_rhoc_correlation}).
    The distribution of critical core densities is directly related to the \ac{CMF}, which shows marginal evidence of a peak consistent with a characteristic post-shock critical mass $M_\mathrm{char,trb}$ associated with the initial cloud-scale Mach number (\cref{fig:mass_function}(a)).
    Resolving the peak of the \ac{CMF} requires the minimum resolvable mass far smaller than the characteristic mass, which becomes increasingly challenging for higher Mach number (Appendix \ref{app:appendix}).

\item 
    Prestellar cores -- i.e. the immediate precursors of stars, as measured at $t_\mathrm{crit}$ -- have distinctive structural and dynamic properties, marking this as a key stage of star formation. 
    However, our analysis implies that the CMF does not map directly to the IMF, at least on the high-mass end.
    In particular, we do not find evidence for massive prestellar cores supported by turbulence.
    Instead, given that the mass accretion rate measured at a fixed radius, $\dot M_\mathrm{in}(r=R_\mathrm{core})$ stays roughly constant 
    or even slightly increases both before and after the collapse (\cref{fig:accretion_rate}(a)), our results suggest that more massive stars may form due to sustained accretion.
    After the sink particle forms, its accretion rate (which in a real system would be the onto a circumstellar disk) remains at a steady level $\sim (10\text{--}20) c_s^3/G$ (\cref{fig:accretion_rate}(b),(d)), comparable to $\dot M_\mathrm{in}(r=R_\mathrm{core})$.
\end{enumerate}

As noted in \citetalias{paperII}, a limitation of the present set of simulations is the absence of magnetic fields.
It will be very interesting to consider models with varying magnetic field strength, in order to test whether the properties of critical cores end up being consistent with the anisotropic model outlined in 
\citet{chen14,chen15}, in the sense of evolving to become magnetically supercritical by inflow along magnetic fields (reaching $v_A\sim c_s$), while simultaneously also having similar turbulence properties (with $\sigma_\mathrm{1D} \sim c_s$) to cores in the present simulations.

\REV{Another caveat is the omission of protostellar feedback in our simulations.
The injection of energy and momentum via radiation and outflows can play a significant role in regulating accretion onto nascent protostars.
Properly modeling these processes is therefore crucial for understanding the full star formation process.
The injected momentum can also locally perturb the velocity field within nearby prestellar cores \citep[e.g.,][]{verliat22,neralwar24}, although protostellar outflows are usually highly collimated such that substantial fraction of core-forming dense gas could remain intact (\citealt{hansen12}; see also \citealt{frank14,bally16} for a review).
In the future, it might be interesting to compare the sonic radius distribution of critical cores with and without outflow feedback included.}

\acknowledgments
We thank the anonymous referee for constructive comments which greatly improved the quality of this paper.
This work was supported in part by grant 510940 from the Simons Foundation to E.~C.\ Ostriker.
Computational resources for this project were provided by Princeton Research Computing, a consortium including PICSciE and OIT at Princeton University.

\bibliographystyle{aasjournal}
\bibliography{mybib}

\appendix

\section{Numerical Resolution in Gravo-Turbulent Simulations}\label{app:appendix}

\begin{deluxetable}{lccccccc}
    \tablecaption{Summary of the simulation parameters in selected works\label{tb:other_works}}
    \tablehead{
\colhead{Model} &
\colhead{$\mathcal{M}_\mathrm{3D}$} &
\colhead{$\Delta x/L_{J,0}$} &
\colhead{$\Delta m/M_{J,0}$} &
\colhead{$M_\mathrm{char,th}/M_{J,0}$} &
\colhead{$M_\mathrm{min}/M_{J,0}$} &
\colhead{$M_\mathrm{min}/M_\mathrm{char,th}$} &
\colhead{$\Delta x(\rho_0)/r_{s,\mathrm{cloud}}$}\\
\colhead{(1)} &
\colhead{(2)} &
\colhead{(3)} &
\colhead{(4)} &
\colhead{(5)} &
\colhead{(6)} &
\colhead{(7)} &
\colhead{(8)}
    }
    \startdata
    \texttt{M5} (this work)                & $5$   & $3.9\times 10^{-3}$ & -                   & $6.6\times 10^{-2}$ & $2.4\times 10^{-2}$ & 0.36  & 0.043 \\ 
    \texttt{M10} (this work)               & $10$  & $3.9\times 10^{-3}$ & -                   & $3.3\times 10^{-2}$ & $2.4\times 10^{-2}$ & 0.72  & 0.087 \\
    \texttt{M2e3\_R3} \citep{guszejnov20}  & $9.3$ & -                   & $5.4\times 10^{-7}$ & $3.6\times 10^{-2}$ & $1.2\times 10^{-3}$ & 0.032 & 0.17 \\
    \texttt{M2e4\_R10} \citep{guszejnov20}  & $16$ & -                   & $1.4\times 10^{-6}$ & $2.1\times 10^{-2}$ & $3.0\times 10^{-3}$ & 0.14 & 0.39 \\
    \texttt{M2e5\_R30} \citep{guszejnov20} & $29$  & -                   & $8.5\times 10^{-6}$ & $1.1\times 10^{-2}$ & $1.8\times 10^{-2}$ & 1.6   & 1.3 \\
    \texttt{high} \citep{haugbolle18}      & $10$  & $3.4\times 10^{-4}$ & -                   & $3.3\times 10^{-2}$ & $2.1\times 10^{-3}$ & 0.063 & 0.35 \\
    \texttt{M8L2N512} \citep{gong15}       & -     & $2.0\times 10^{-3}$ & -                   & $8.6\times 10^{-2}$ & $1.2\times 10^{-2}$ & 0.14  & -
    \enddata
    \tablecomments{Columns (3) and (4) provide the highest resolution (i.e., smallest $\Delta x$ for Eulerian simulations and smallest $\Delta m$ for Lagrangian simulations) of the models used for convergence study given in Column (1). All simulations are inherently scale-free and the choice of $L_{J,0}$ and $M_{J,0}$ is arbitrary. Because \citet{guszejnov20} and \citet{haugbolle18} present their results in physical units, we provide their adopted values for convenience: $L_{J,0} = 1.3\,\mathrm{pc}$, $M_{J,0} = 37\,M_\odot$ (\texttt{M2e3\_R3}); $L_{J,0} = 2.4\,\mathrm{pc}$, $M_\mathrm{J,0}=71\,M_\odot$ (\texttt{M2e4\_R10}); $L_{J,0} = 4.0\,\mathrm{pc}$, $M_\mathrm{J,0}=120\,M_\odot$ (\texttt{M2e5\_R30}); $L_{J,0} = 0.71\,\mathrm{pc}$, $M_{J,0} = 17\,M_\odot$ (\texttt{high}).
    Because the inflow Mach numbers in \citet{gong15} cannot be directly related to $\mathcal{M}_\mathrm{3D}$, we use the directly measured post-shock density to calculate $M_\mathrm{char,th}$ rather than from \cref{eq:mchar} (see text).
    For the same reason, it is not straightforward to infer $r_{s,\mathrm{cloud}}$ in their simulations; we thus leave it out in this table.}
\end{deluxetable}

To cross-compare the numerical resolutions of different simulations that have been used to address the question of numerical convergence of mass functions, \cref{tb:other_works} lists resolution parameters in some recent simulations, including this work.
Column (1) gives the model name and reference.
Column (2) gives the cloud-scale Mach number.
Columns (3) and (4) give the numerical resolution $\Delta x$ and $\Delta m$ for fixed spatial- and fixed mass-resolution simulations, respectively, normalized by the Jeans scales at the mean cloud condition (for the \ac{AMR} simulations, we indicate $\Delta x$ at the highest refinement level).
Column (5) gives the characteristic mass $M_\mathrm{char,th}$ from \cref{eq:mchar}.
Column (6) gives the minimum resolvable mass $M_\mathrm{min}$ from \cref{eq:Mmin_given_dx,eq:Mmin_given_dm}, adopting $N_\mathrm{core,res} = 8$ and using either $\Delta x$ or $\Delta m$ given in Columns (3)--(4).
Column (7) gives the ratio $M_\mathrm{min}/M_\mathrm{char,th}$.
Column (8) gives the numerical resolution at the mean density divided by the sonic scale, where we take the effective spatial resolution $\Delta x(\rho_0) = (\Delta m/\rho_0)^{1/3}$ for fixed mass-resolution simulations.

If the core mass function peaks roughly around $M_\mathrm{char,th}$ predicted by various theories, clearly resolving the peak in numerical simulations would require $M_\mathrm{min}/M_\mathrm{char,th} \ll 1$.
Combining \cref{eq:Mmin_given_dx,eq:Mmin_given_dm,eq:mchar}, one can show that
\begin{subequations}\label{eq:Mmin_over_Mchar}
\begin{align}
    \frac{M_\mathrm{min}}{M_\mathrm{char,th}} &=    
        2.32 \mathcal{M}_\mathrm{3D} N_\mathrm{core,res}\left(\frac{\Delta x}{L_{J,0}}\right)\\
        &= 12.5 \mathcal{M}_\mathrm{3D} N_\mathrm{core,res}^3 \left(\frac{\Delta m}{M_{J,0}}\right).
\end{align}
\end{subequations}
This ratio is linearly proportional to the resolution (either $\Delta x$ or $\Delta m$ normalized to the Jeans scales at mean density), and is also linear in the Mach number, making it more difficult to resolve the peak of the \ac{CMF} in higher Mach number simulations. 
Although \cref{tb:other_works} is far from being complete, \cref{eq:Mmin_over_Mchar} suggests that simulations should typically find the peak of the \ac{CMF} (and possibly the \ac{SMF}, modulo disk fragmentation) converges provided that ${M_\mathrm{min}}/{M_\mathrm{char,th}}$ is sufficiently small.
For our lower Mach number simulations, the ratio $M_\mathrm{min}/M_\mathrm{char,th}=0.36$, while for our higher Mach number simulations this ratio is 0.72. Thus, we expect to identify the \ac{CMF} peak clearly in our {\tt M5} model but not in our {\tt M10}, and indeed this appears consistent with our results.

The model \citet{guszejnov20} used for their \ac{SMF} numerical convergence study (\texttt{M2e5\_R30}) has a fairly high Mach number $\mathcal{M}_\mathrm{3D}=29$, such that $M_\mathrm{min}/M_\mathrm{char,th}=1.6$  at their highest mass resolution (i.e., smallest $\Delta m = 10^{-3}\,M_\odot = 8.5\times 10^{-6}M_{J,0}$.  Thus, this set of numerical parameters would not be suitable for identifying the peak of \ac{CMF}.  However, the lower Mach number models in \citet{guszejnov20} (\texttt{M2e3\_R3} and \texttt{M2e4\_R10}) have ratio $M_\mathrm{min}/M_\mathrm{char,th} = 0.032\text{--}0.14$, making them suitable to resolve a \ac{CMF} peak.
Indeed, their rerun of model \texttt{M2e4\_R10}, after the error in the sink particle algorithm was corrected \citep{guszejnov21}, shows a peak in the \ac{SMF} near $M_\mathrm{char,th} = 2.1\times 10^{-2}M_{J,0} = 1.5\,M_\odot$ (conversion to physical units adopting their sound speed and mean density).

In the \ac{AMR} simulations of a $\mathcal{M}_\mathrm{3D} = 10$ cloud conducted by \citet{haugbolle18}, the minimum resolvable mass at their highest refinement level ($\Delta x = 50\,\mathrm{au} = 3.4\times 10^{-4}L_{J,0}$ adopting their mean cloud density and sound speed) is $M_\mathrm{min} = 2.1\times 10^{-3} M_{J,0}(N_\mathrm{core,res}/8)$ from \cref{eq:Mmin_given_dx}.
This leads to $M_\mathrm{min}/M_\mathrm{char,th}=0.063$ if we adopt $N_\mathrm{core,res} = 8$.
\citet{pelkonen21} performed convergence analysis using the same simulation data of \citet{haugbolle18}, finding that the \ac{SMF} in their simulation converges to $0.17\,M_\odot$ with increasing numerical resolution.
They also found that the corresponding progenitor core mass is $0.26\,M_\odot = 1.5\times 10^{-2}M_{J,0}$, which is a factor of $2.2$ lower than $M_\mathrm{char,th}$.
We note that the \ac{CCMF} in our model \texttt{M5} also suggests a peak at a factor of $\sim 2$ lower than $M_\mathrm{char,th}$ (\cref{fig:mass_function}(a)), similar to the results of \citet{pelkonen21}.

\citet{gong15} performed local simulations focusing on a small portion of a \ac{GMC} where the flows are supersonically converging, using an idealized converging flow setup.
Due to the instability in the non-magnetized post-shock layer, their measured average density within the post-shock layer is only weakly dependent on the \emph{inflow} Mach number such that \cref{eq:mchar} is not straightforwardly applicable.
Instead, we use their measured average post-shock density $\rho_\mathrm{ps} \sim 15\rho_0$ to estimate the characteristic mass $M_\mathrm{char,th}=M_\mathrm{BE}(\overline{\rho}=\rho_\mathrm{ps}) = 8.6\times 10^{-2}M_{J,0}$.
The minimum resolvable mass at their highest resolution model with $\Delta x = 2\times 10^{-3}L_{J,0}$ is $M_\mathrm{min} = 1.2\times 10^{-2}M_{J,0}(N_\mathrm{core,res}/8)$, a factor of $7$ lower than $M_\mathrm{char,th}$ for $N_\mathrm{core,res}=8$.
They found that the peak of the \ac{CMF} is numerically converged (see their Figure 9), with the measured peak location consistent with $M_\mathrm{char,th}$.
It is worth noting that \citet{gong15} used the \ac{CMF} constructed from the core masses measured at the end of collapse ($t_\mathrm{coll}$) for individual core.

Although we choose the resolution of our simulations carefully to resolve most of the core formation \emph{by mass} (see Section \secresolution of \citetalias{paperII}), resolving the peak of the \ac{CMF} poses a severe challenge because the mass fraction below the peak is very small.
For example, if we assume the peak of the \ac{CMF} occurs at $M_\mathrm{peak}\sim 0.5 M_\mathrm{char,th}$ as it appears to be the case in this work and in \citet{pelkonen21}, clear identification of the peak would require at least $M_\mathrm{min} < (1/2) M_\mathrm{peak} = 0.25M_\mathrm{char,th}$.
\cref{eq:Mmin_over_Mchar} then leads to the following resolution requirement for simulations with a fixed spatial resolution:
\begin{equation}\label{eq:dx_req_imf}
    \frac{\Delta x}{L_{J,0}} < 1.35\times 10^{-3} \left(\frac{N_\mathrm{core,res}}{8}\right)^{-1}\left(\frac{\mathcal{M}_\mathrm{3D}}{10}\right)^{-1},
\end{equation}
and the requirement for fixed mass resolution:
\begin{equation}\label{eq:dm_req_imf}
    \frac{\Delta m}{M_{J,0}} < 3.91\times 10^{-6} \left(\frac{N_\mathrm{core,res}}{8}\right)^{-3}\left(\frac{\mathcal{M}_\mathrm{3D}}{10}\right)^{-1}.
\end{equation}
Considering that the size of the computational domain should be properly scaled with $\mathcal{M}_\mathrm{3D}$ to keep the cloud scale virial parameter constant, the total number of simulation elements for a uniform mesh would be
\begin{equation}\label{eq:nreq_imf_eul}
\begin{split}
    N^3 &= \left(\frac{L_\mathrm{box}}{\Delta x}\right)^3\\
      &= (2698)^3 \left(\frac{N_\mathrm{core,res}}{8}\right)^3\left(\frac{\alpha_\mathrm{vir,box}}{2}\right)^{-3/2}\left(\frac{\mathcal{M}_\mathrm{3D}}{10}\right)^6,
\end{split}
\end{equation}
where we used \cref{eq:cloud_virial_parameter}.
Similarly, the required number of resolution elements for the fixed mass-resolution simulations is
\begin{equation}\label{eq:nreq_imf_lag}
\begin{split}
    N_p &\equiv \frac{M_\mathrm{box}}{\Delta m}\\
        &= 1.24\times 10^7\left(\frac{N_\mathrm{core,res}}{8}\right)^3\left(\frac{\alpha_\mathrm{vir,box}}{2}\right)^{-3/2}\left(\frac{\mathcal{M}_\mathrm{3D}}{10}\right)^4.    
\end{split}
\end{equation}
We note that \cref{eq:nreq_imf_eul,eq:nreq_imf_lag} are minimum requirements and that even higher resolutions would be required to robustly demonstrate the numerical convergence of the identified peak.

Taken at face value, \cref{eq:nreq_imf_eul,eq:nreq_imf_lag} suggest that it might be easier to resolve the peak of the \ac{CMF} by adopting a fixed mass-resolution rather than fixed spatial resolution.
However, it is worth emphasizing that in addition to \cref{eq:dx_req_imf} for resolving cores, the effective spatial resolution in low-density regions must also be sufficiently high to resolve the sonic scale of turbulence.
Column (8) of \cref{tb:other_works} shows that, compared to other current  simulations with adaptive resolution, the uniform resolution adopted in this work allows smaller $\Delta x(\rho_0)/r_{s,\mathrm{cloud}}$ and thus better resolves early nonlinear structures created by turbulence.

More generally, \citetalias{paperII} (see Equation \eqnminsonic there) shows that resolving $r_{s,\mathrm{cloud}}$ imposes a resolution requirement that has the same Mach number scaling as \cref{eq:nreq_imf_eul}.
However, \citetalias{paperII} (see  Section \secresolutiondiscussion there) also demonstrated that the local sonic scale exhibits an order of magnitude variations (which is correlated with density) above and below $r_{s,\mathrm{cloud}}$, which then demands higher resolution in moderate-density gas.
To meet this additional requirement at mean cloud density, i.e., $\Delta x(\rho_0) \ll r_{s,\mathrm{cloud}}$, simulations adopting a fixed mass resolution need to satisfy not only \cref{eq:nreq_imf_lag} but also additionally
\begin{equation}
    N_p = \frac{M_\mathrm{box}}{\Delta m} = 1.90\times 10^{10}\left(\frac{N_{s,\mathrm{res}}}{30}\right)^3 \left(\frac{\mathcal{M}_\mathrm{3D}}{10}\right)^6,
\end{equation}
where $N_{s,\mathrm{res}} = r_{s,\mathrm{cloud}}/\Delta x(\rho_0) = 30$ is recommended in \citetalias{paperII}.\footnote{In \citetalias{paperII}, we defined $N_{s,\mathrm{res}}$ with respect to sonic scale $\lambda_{s,\mathrm{cloud}} = (4/3)r_{s,\mathrm{cloud}}$ and recommended $\lambda_{s,\mathrm{cloud}} / \Delta x = 40$, which is equivalent to $N_{s,\mathrm{res}} = 30$ presented here.}
Thus, properly resolving the sonic scale is in fact more computationally challenging than resolving self-gravitating cores, whether for fixed mesh or moving mesh simulations.

\end{document}